\documentclass[preprint,showpacs,amsmath,amssymb,aps]{revtex4}
\usepackage{graphicx}
\begin{document}

\title{ A new picture for the chiral symmetry properties within a particle-core framework}

\author{A. A. Raduta$^{a),b)}$\footnote{Senior Fellow of Humboldt Foundation},  C. M. Raduta$^{a)}$ and  Amand Faessler$^{c)}$   }

\address{$^{a)}$ Department of Theoretical Physics, Institute of Physics and
  Nuclear Engineering, Bucharest, POBox MG6, Romania}

\address{$^{b)}$Academy of Romanian Scientists, 54 Splaiul Independentei, Bucharest 050094, Romania}

\address{$^{c}$Institut f\"{u}r Theoretische Physik der Universit\"{a}t  
T\"{u}bingen, D-72076 T\"{u}bingen, Germany}

\begin{abstract}
The Generalized Coherent State Model, proposed previously for a unified description of magnetic and electric collective properties of nuclear systems, is extended to account for the chiral like properties of nuclear systems. To a phenomenological core described by the GCSM a set of interacting particles are coupled. Among the particle-core states one identifies a finite set which have the property that the angular momenta carried by the proton and neutron quadrupole bosons and the particles respectively, are mutually orthogonal. All terms of the model Hamiltonian satisfy the chiral symmetry except for the spin-spin interaction. The magnetic properties of the 
particle-core states, where the three mentioned angular momenta are orthogonal, are studied. A quantitative comparison of these features with the similar properties of states, where the three angular momenta belong to the same plane, is performed. 
\end{abstract}


\maketitle

\renewcommand{\theequation}{1.\arabic{equation}}
\setcounter{equation}{0}

\section{Introduction}
\label{sec:level 1}
The rotational spectra appear to be a reflection of a spontaneous  rotational symmetry breaking when the nuclear system acquires a static nuclear deformation. The fundamental nuclear properties  like nuclear shape, the nuclear  mass and charge distribution inside the nucleus, electric and magnetic moments, collective spectra may be evidenced through the system interaction with an electromagnetic field. The two components of the field, electric and magnetic, are used to explore the properties of electric and magnetic nature, respectively. At the end of last century the scissors like states \cite{LoIu1,LoIu2} as well as the spin-flip excitations \cite{Zawischa} have been widely treated by various groups. Some of them were based on phenomenological assumptions while the other ones on microscopic considerations. The scissors like excitations are excited  in (e,e') experiments at backward angles and expected at an energy of about 2-3 MeV, while the spin-flip excitations are seen in (p,p') experiments at forward angles and are located at about 5-10 MeV. The scissors mode describes the angular oscillation of proton against neutron system and the total strength is proportional to the nuclear deformation squared which reflects the collective character of the excitation. Many papers have been written on this subject and therefore it is difficult to quote all of them. We mention however two reviews given in Refs. \cite{LoIu3,Zawischa}.

Since the total M1 strength of the $M1$ mode is proportional to the nuclear deformation squared, it was believed that the magnetic collective properties are in general associated with deformed systems. This is not true due to the magnetic dipole bands, where the ratio between the moment of inertia and the B(E2) value for exciting the first $2^+$ from the ground state $0^+$,
${\cal I}^{(2)}/B(E2)$,
takes large values, of the order of 100(eb)$^{-2}MeV^{-1}$. These large values can be justified
by a large transverse magnetic dipole moment (perpendicular to the total angular momentum) which induces dipole magnetic transitions, but almost no charge quadrupole moment \cite{Frau}. Indeed, there are several experimental data showing that the dipole bands have large values for $B(M1)\sim 3-6\mu^2_N$ and very small values of $B(E2)\sim 0.1(eb)^2$ (see for example Ref.\cite{Jenkins}). The states are different from the scissors mode, they being rather of a shears character. A system with a large transverse magnetic dipole moment (the component of the magnetic moment perpendicular to the total angular momentum)  which was studied in many publications, may consist of a triaxial core to which a proton prolate  and a  neutron oblate hole orbital are coupled. The interaction of particle and hole like orbitals is repulsive, which keeps the two orbits apart from each other. In this way the orthogonal angular momenta carried by the proton particles and neutron holes are favored. The maximal transverse dipole momentum is achieved, for example, when ${\bf j}_p$ is oriented along the small axis of the core, ${\bf j}_n$ along the long axis and the core rotates around the intermediate axis.   Suppose the three orthogonal angular momenta form a right trihedral frame. If the Hamiltonian describing the interacting system of protons, neutrons and the triaxial core is invariant to the transformation which changes the orientation of one of the three angular momenta, i.e. the right trihedral frame is transformed to a left type, one says that the system exhibits a chiral symmetry. As always happens, such a symmetry is identified when that is broken and consequently to the two trihedrals correspond distinct energies, otherwise close to each other. Thus, a signature for a chiral symmetry characterizing a triaxial system is the existence of two $\Delta I=1$ bands which are close in energies.  Increasing the total angular momentum the gradual alignment of ${\bf j}_p$ and ${\bf j}_n$ to the total ${\bf J}$ takes place and a magnetic band is developed.

{\it{The question addressed in this paper is whether the picture of the three angular momenta system, carried by a phenomenological core, a prolate and an oblate single particle orbitals, with respect to which the chiral symmetry is defined is unique for determining states connected with large M1 transitions. Note that the nuclear system which accommodate the chiral frame is odd-odd.}}

In the past, the magnetic states of orbital or of spin-flip nature were considered by our group in several publications \cite{1,2,3,4,5,6,7,8,9,10}.
We studied  also the dipole bands with  $K^{\pi}=1^{\pm}$ using a quadrupole and octupole boson Hamiltonian and a set of model states obtained by parity and angular momentum projections from a quadrupole deformed ground state without space reflection symmetry \cite{11}. We pointed out that the band $1^+$ has a magnetic character while the dipole band $1^-$ is of an electric type.
In another publication \cite{12} we pointed out that the parity partner bands have the property that starting from a critical angular momentum, the states have the property that the angular momenta carried by the quadrupole and octupole bosons respectively, are mutually orthogonal. Therefore one may expect that adding to the phenomenological Hamiltonian a set of interacting particles one could achieve a configuration where the angular momentum carried by nucleons is perpendicular on the quadrupole and octupole  angular momenta which are already orthogonal. The first attempt was already made in Ref.\cite{13}.  

Here we attempt another chiral system consisting of one phenomenological core with two components, one for protons and one for neutrons, and two quasiparticles whose total angular momentum is oriented along the symmetry axis of the core due to the particle-core interaction. We investigate whether
states of total angular momentum ${\bf I}$, where the three components mentioned above carry angular momenta, ${\bf J}_p, {\bf J}_n, {\bf J}_F$, which are mutually orthogonal, may exist. We believe that if such configuration exists it is optimal for  defining large transverse magnetic moment inducing large M1 transitions. 

\renewcommand{\theequation}{2.\arabic{equation}}
\setcounter{equation}{0}

\section{The Generalized Coherent State Model}
\label{sec:level 2}
The description of magnetic properties in nuclei has always been a central
issue. The reason is that the two systems of protons and neutrons respond
differently when they interact with an external electromagnetic field.
Differences are due to the fact that by contrast to neutrons,  protons are
charged particles, the proton and neutron magnetic moments are different from
each other and, finally, the proton and neutron numbers in a given nucleus are, in general,  different.

Many papers have been devoted to explaining  various features of the collective
dipole mode called, conventionally, scissors mode. The name of the mode  was suggested by Lo Iudice and Palumbo
who interpreted the dipole mode, within the Two Rotor Model \cite{LoIu1}, as a scissors like oscillation of proton and
neutron systems described by two axially symmetric ellipsoids, respectively.

The Coherent State Model (CSM), proposed by Raduta {\it et al.} to describe the lowest three collective interacting bands \cite{Rad1},  was extended  by including the isospin degrees of freedom in order to
 account for the collective  properties of the scissors mode \cite{Rad2}.
This extension is conventionally called ``The Generalized Coherent State Model''(GCSM).

CSM starts with the construction of a restricted collective space, by projecting out the components of good angular momentum from three orthogonal quadrupole boson states. These states are chosen such that they are orthogonal before and after 
projection.
One of the three deformed states, the intrinsic ground state, is a coherent state of Glauber type with respect to the zero component of the quadrupole boson, $b^{\dagger}_{20}$, while the other two are obtained by acting with elementary boson polynomials on the ground state. In choosing the intrinsic excited states we take care that the projected states considered 
 in the vibrational limit have to provide the multi-phonon vibrational spectrum,  while for the large deformation regime their behavior coincides with that predicted by the liquid drop model.   
 
In contrast to the CSM, which uses only one boson for the composite system
of protons and neutrons, within the GCSM the protons are described by quadrupole
proton-like bosons, $b^{\dagger}_{p\mu}$, while the neutrons by quadrupole neutron-like bosons, $b^{\dagger}_{n\mu}$ .
Since one deals with two quadrupole bosons instead of one, one expects 
to have a more flexible model and to find a simpler solution satisfying the restrictions
required by CSM.  The restricted  collective space is defined  by the states describing the three
major bands, ground, beta and gamma, as well as the band  based on
the  isovector state $1^+$. Orthogonality conditions, required for both intrinsic and projected states, are satisfied by the
following 6 functions which generate by angular momentum projection,
6 rotational bands:
\begin{eqnarray}
\phi^{(g)}_{JM}&=&N^{(g)}_JP^J_{M0}\Psi_g,~~
\Psi_g=exp[(d_pb^{\dag}_{p0}+d_nb^{\dag}_{n0})-(d_pb_{p0}+d_nb_{n0})]
|0\rangle,
\nonumber\\
\phi^{(\beta)}_{JM}&=&N^{(\beta)}_JP^J_{M0}\Omega_{\beta}\Psi_g,
\nonumber\\
\phi^{(\gamma)}_{JM}&=&N^{(\gamma)}_JP^J_{M2}(b^{\dag}_{n2}-b^{\dag}_{p2})\Psi_g,
\nonumber\\
\tilde{\phi}^{(\gamma)}_{JM}&=&\tilde{N}^{(\gamma)}_JP^J_{M2}(\Omega^{\dag}_{\gamma,p,2}+\Omega^{\dag}_{\gamma,n,2})\Psi_g,
\nonumber\\
\phi^{(1)}_{JM}&=&N^{(1)}_JP^J_{M1}(b^{\dag}_nb^{\dag}_p)_{11}\Psi_g,
\nonumber\\
\tilde{\phi}^{(1)}_{JM}&=&\tilde{N}^{(1)}_JP^J_{M1}(b^{\dag}_{n1}-b^{\dag}_{p1})\Omega^{\dag}_{\beta}\Psi_g.
\label{figcsm}
\end{eqnarray}
Here, the following notations have been used:
\begin{eqnarray}
\Omega^{\dag}_{\gamma,k,2}&=&(b^{\dag}_kb^{\dag}_k)_{22}+d_k\sqrt{\frac{2}{7}}
b^{\dag}_{k2},~~k=p,n,
\nonumber\\
\Omega^{\dag}_{\beta}&=&\Omega^{\dag}_p+\Omega^{\dag}_n-2\Omega^{\dag}_{pn},
\nonumber\\
\Omega^{\dag}_k&=&(b^{\dag}_kb^{\dag}_k)_0-\sqrt{\frac{1}{5}}d^2_k,~~k=p,n,
\nonumber\\
\Omega^{\dag}_{pn}&=&(b^{\dag}_pb^{\dag}_n)_0-\sqrt{\frac{1}{5}}d^2_p.
\label{omegagen}
\end{eqnarray}
Note that a-priory we cannot select one of the two sets of states
$\phi^{(\gamma)}_{JM}$ and $\tilde{\phi}^{(\gamma)}_{JM}$ for gamma band, although  one is symmetric and the other asymmetric against proton-neutron permutation.
The same is true for the two isovector candidates for the dipole states.
In Ref.\cite{Rad3}, results obtained by using alternatively a symmetric and an asymmetric structure
for the gamma band states were presented. Therein it was shown that the asymmetric structure
for the gamma band does not conflict any of the available data. By contrary,
considering for the gamma states an asymmetric structure and fitting the model
Hamiltonian coefficients in the manner described  in Ref.\cite{Rad3}, a better
description for the beta band energies is obtained. Moreover, in that situation
the description of the E2 transition becomes technically very simple.
For these reasons, here we make the option for a proton neutron asymmetric
gamma band.

All calculations performed so far considered equal deformations for protons and neutrons. The deformation parameter for the composite system is:
\begin{equation}
\rho=\sqrt{2}d_p=\sqrt{2}d_n \equiv \sqrt{2}d.
\end{equation}
The factors $N$ involved in the wave functions are normalization constants calculated in terms of some overlap integrals.

We seek now an effective Hamiltonian for which the projected states (\ref{figcsm}) are, at least in a good approximation, eigenstates in the restricted collective space.
The simplest Hamiltonian fulfilling this condition is:
\begin{eqnarray}
H_{GCSM}&=&A_1(\hat{N}_p+\hat{N}_n)+A_2(\hat{N}_{pn}+\hat{N}_{np})+
\frac{\sqrt{5}}{2}(A_1+A_2)(\Omega^{\dag}_{pn}+\Omega_{np})
\nonumber\\
&&+A_3(\Omega^{\dag}_p\Omega_n+\Omega^{\dag}_n\Omega_p-2\Omega^{\dag}_{pn}
\Omega_{np})+A_4\hat{J}^2.
\label{HGCSM}
\end{eqnarray}
Here $\hat{N}_i$ with $i=p, n, pn$ denotes the boson number operators:
\begin{equation}
\hat{N}_{pn}=\sum_{m}b^{\dag}_{pm}b_{nm},~\hat{N}_{np}=(\hat{N}_{pn})^{\dag},~~
\hat{N}_k=\sum_{m}b^{\dag}_{km}b_{km},~k=p,n.
\end{equation}
The Hamiltonian given by Eq.(\ref{HGCSM}) has  only one off-diagonal matrix element in the basis (\ref{figcsm}). That is $\langle\phi^{\beta}_{JM}|H|\tilde{\phi}^{(\gamma)}_{JM}\rangle$.
However, our calculations show that this affects the energies of $\beta$ and $\tilde{\gamma}$ bands by an amount of a few keV. Therefore, the excitation energies of the six bands, are in a very good approximation, given by the diagonal element:
\begin{equation}
E^{(k)}_J=\langle\phi^{(k)}_{JM}|H|\phi^{(k)}_{JM}\rangle-
\langle\phi^{(g)}_{00}|H|\phi^{(g)}_{00}\rangle,\;\;k=g,\beta,\gamma,1,\tilde{\gamma},\tilde{1}.
\label{EkJ}
\end{equation}
It can be easily checked that the model Hamiltonian does not commute with the components of the $\hat{F}$ spin operator:
\begin{equation}
\hat{F}_0=\frac{1}{2}(\hat{N}_p-\hat{N}_n),\;\hat{F}_+=\hat{N}_{pn},\;\hat{F}_-=\hat{N}_{np}.
\label{Fspin}
\end{equation}

Hence, the eigenstates of $H$ are $F_0$ mixed states. However, the expectation
values of the $F_0$ operator on the projected model states are equal to zero.
This is caused by the fact that the proton and neutron deformations are considered
to be equal. In this case the states are of definite parity with respect to
the proton-neutron permutation, which is consistent with the structure of the
model Hamiltonian which is invariant with respect to such a symmetry
transformation. To conclude, by contrast to the IBA2 Hamiltonian, the GCSM
Hamiltonian is not $\hat{F}$ spin invariant. Another difference to the IBA2, the most
essential one, is that the GCSM Hamiltonian does not commute with the boson number
operators. Due to this feature the coherent state approach proves to be the
most adequate one to treat the Hamiltonian in Eq.(2.4).
The asymptotic behavior of the magnetic state $1^+$, derived in Ref.\cite{Rad2},
shows clearly that
the phenomenological description of two liquid drops and two rigid rotors are
just particular cases  of the GCSM, defined by specific restrictions.

The GCSM seems to be the only phenomenological model which treats simultaneously the
M1 and E2 properties. Indeed, in Refs.\cite{Rad3,Rad4} the ground, beta and gamma bands
are considered
together with a $K^{\pi}=1^+$ band built on the top of the scissor mode $1^+$.
By contrast to the other phenomenological and microscopic models, which treat
the scissors mode in the intrinsic reference frame, here one deals with states of good
angular momentum and, therefore,
there is no need to restore the rotational symmetry.
As shown in Ref.\cite{Iud} the GCSM provides
for the total M1 strength an expression which is proportional to the
nuclear deformation
squared. Consequently, the M1 strength of $1^+$ and the B(E2) value
for $2^+$ are
proportional to each other, although the first quantity is determined by the
convection current
while the second one by the static charge distribution.

One weak point of most phenomenological models is that they use expressions
for transition operators not consistent with the structure of the
model Hamiltonian. Thus, the transition probabilities are influenced by the
chosen Hamiltonian only through the wave functions.
By contradistinction in Refs. \cite{Rad3,Rad4} the E2 transition operator, as well as the M1 form-factor, are derived analytically, by
using the equation of motion of the collective coordinates determined by the model Hamiltonian. In this way a consistent description of electric and magnetic properties of many nuclei was attained.

\section{Proton and neutron angular momenta composition of the ground and dipole magnetic bands}
\renewcommand{\theequation}{3.\arabic{equation}}
\setcounter{equation}{0}

We start by mentioning few properties for the intrinsic ground state wave function, $\Psi_g$. Note that $\Psi_g$ can be written in a factorize form:
\begin{equation}
\Psi_{g} \equiv \Psi_p\Psi_n,
\end{equation}
where the factor functions are:
\begin{equation}
\Psi_p=exp[d_pb^{\dag}_{p0}-d_pb_{p0}]|0\rangle_p,\;\; \Psi_n=exp[d_nb^{\dag}_{n0}-d_nb_{n0}]|0\rangle_n.
\end{equation}
The $\tau$ functions, with $\tau=p,n$, are eigenstates of the z projection of the angular momentum and therefore can be expanded in the basis $|J_{\tau}0\rangle$ defined by the eigenstates of 
${\bf J}_{\tau}^2,J_{\tau 0}$:
\begin{equation}
\Psi_{\tau}=\sum_{J_{\tau}}C_{J_{\tau}}|J_{\tau}0\rangle, \tau =p,n.
\label{expans}
\end{equation}
Denoting by
\begin{equation}
\varphi^{(g)}_{J_{\tau}M_{\tau}}= N^{(g)}_{J_{\tau}}P^{J_{\tau}}_{M_{\tau}0}\Psi_{\tau},
\label{prgrst}
\end{equation}
the angular momentum projected state associated to $\Psi_{\tau}$ and then inserting the expression (\ref{expans}) in the right hand side of (\ref{prgrst}), one finds that the expansion coefficients $C_{J_{\tau}}$ are related with the projected state norms, by:
\begin{equation}
 C_{J_{\tau}}=\left(N^{(g)}_{J_{\tau}}\right)^{-1}.
\end{equation}
Here $N^{(g)}_{J_p}$ and $N^{(g)}_{J_n}$ denote the norms of the angular momentum projected states associated to $\Psi_p$ and $\Psi_n$, respectively. These have been analytically expressed in 
Ref.\cite{Rad1}, where the projected states $\varphi^{(g)}_{J_{\tau}M_{\tau}}$ are used as model states for the rotational ground band.

The above analysis can be easily extended to the intrinsic ground state describing the composite proton-neutron system:
\begin{equation}
\Psi_{g}=\Psi_p\Psi_n=\sum_{J_p,J_n=even}C_{J_p}|J_p0\rangle C_{J_n}|J_n0\rangle
        =\sum_{J_p,J_n,J}C_{J_p}C_{J_n}C^{J_p\;J_n\;J}_{0\;\;0\;\;0}|J,0\rangle .  
\end{equation}
The angular momentum projected state is defined by:
\begin{eqnarray}
\phi^{(g)}_{JM}&=&N^{(g)}_{J}P^J_{M0}\Psi_{g}=N^{(g)}_{J}\sum_{J_pJ_n}C_{J_p}C_{J_n}C^{J_p\;J_n\;J}_{0\;\;0\;\;0}|J,M\rangle\nonumber\\
&=&N^{(g)}_{J}\sum_{J_pJ_n}\left(N^{(g)}_{J_p}\right)^{-1}\left(N^{(g)}_{J_n}\right)^{-1}
C^{J_p\;J_n\;J}_{0\;\;0\;\;0}\left[\varphi^{(g)}_{J_p}\varphi^{(g)}_{J_n}\right]_{JM},
\end{eqnarray}
with the norm:
\begin{equation}
\left(N^{(g)}_{J}\right)^{-2}=
\sum_{J_p,J_n}\left(N^{(g)}_{J_p}\right)^{-2}\left(N^{(g)}_{J_n}\right)^{-2}\left(C^{J_p\;J_n\;J}_{0\;\;0\;\;0}\right)^{2} .
\end{equation}
In the above equations the standard notation for the Clebsch-Gordan coefficients has been used.

The average value of the angular momentum carried by the proton bosons is given by:
\begin{equation}
\langle \phi^{(g)}_{JM}|{\bf \hat{J}}_p^2|\phi^{(g)}_{JM}\rangle =
\left(N^{(g)}_J\right)^{2}\sum_{J_p,J_n}\left(N^{(g)}_{J_p}\right)^{-2}\left(N^{(g)}_{J_n}\right)^{-2}J_p(J_p+1)\left(C^{J_p\;J_n\;J}_{0\;\;0\;\;0}\right)^2\equiv 
\widetilde{J}^{(g)}_{pJ}(\widetilde{J}^{(g)}_{pJ}+1).
\end{equation}
Similarly, one calculates the average angular momentum carried by the neutron bosons, 
$\widetilde{J}^{(g)}_{nJ}$. The two angular momenta, $\widetilde{J}^{(g)}_{pJ}$, $\widetilde{J}^{(g)}_{nJ}$, define the relative angle which obey the equation:
\begin{equation}
\cos({\bf J}_p,{\bf J}_n)^{(g)}_J=\frac{J(J+1)-\widetilde{J}^{(g)}_{pJ}(\widetilde{J}^{(g)}_{pJ}+1)-\widetilde{J}^{(g)}_{nJ}(\widetilde{J}^{(g)}_{nJ}+1)}{2\sqrt{\widetilde{J}^{(g)}_{pJ}(\widetilde{J}^{(g)}_{pJ}+1)\widetilde{J}^{(g)}_{nJ}(\widetilde{J}^{(g)}_{nJ}+1)}}.
\end{equation}

Let us consider now the angular momentum projection of following dipole excitation of the intrinsic ground state
\begin{eqnarray}
\phi^{(1)}_{JM}&=&N^{(1)}_JP^J_{M1}(b^{\dag}_nb^{\dag}_p)_{11}\psi_g\nonumber\\
               &=&N^{(1)}_J\sum_{J'=even}\left(N^{(g)}_{J'}\right)^{-1}C^{J'\;1\;J}_{0\;\;1\;\;1}\left[\left(b^{\dag}_n
b^{\dag}_p\right)_{1}\varphi^{(g)}_{J'}\right]_{JM},
\end{eqnarray}
with the norm having the expression:
\begin{equation}
\left(N^{(1)}_J\right)^{-2}=\sum_{J'=even}\left(N^{(g)}_{J'}\right)^{-2}
\left(C^{J'\;1\;J}_{0\;\;1\;\;1}\right)^2 .
\end{equation}

It is worth calculating the separate contributions of  proton and  neutron bosons to building up the total angular momentum of a given magnetic dipole state. The effective angular momentum $\tilde{J}$ is defined as:
\begin{eqnarray}
&&\widetilde{J}^{(1)}_{p;J}(\widetilde{J}^{(1)}_{p;J}+1)=\langle\phi^{(1)}_{JM}|{\bf \hat{J}}_p^2|\phi^{(1)}_{JM}\rangle
\nonumber\\
&=&6+\left(N^{(1)}_J\right)^{2}\sum_{J_p,J_n,J'}\left(N^{(g)}_{J_p}\right)^{-2}\left(N^{(g)}_{J_n}\right)^{-2}J_p(J_p+1)\left(C^{J_p\;J_n\;J'}_{0\;\;0\;\;0}\right)^2
\left(C^{J'\;1\;J}_{0\;\;1\;\;1}\right)^2 .
\end{eqnarray}
Since the ground state is symmetric with respect to the $p-n$ permutation, one expects that the effective neutron angular momentum defined by averaging the operator $\hat{J}_{n;J}^2$
with the ground state projected function is equal to the effective proton angular momentum, i.e.
\begin{equation}
\widetilde{J}^{(1)}_{n;J}=\widetilde{J}^{(1)}_{p;J}.
\end{equation}
Denoting the ground state angular momentum by
\begin{equation}
{\bf J}_{pn}={\bf J}_p+{\bf J}_n,
\label{defJ'}
\end{equation}
then for the average value one obtains:
\begin{equation}
\widetilde{J}^{(1)}_{pn;J}(\widetilde{J}^{(1)}_{pn;J}+1)\equiv\langle\phi^{(1)}_{JM}|{\bf \hat{J}}_{pn}^{2}|\phi^{(1)}_{JM}\rangle
=\left(N^{(1)}_J\right)^{2}\sum_{J''}\left(N^{(g)}_{J''}\right)^{-2}\left(C^{J''\;1\;J}_{0\;\;1\;\;1}\right)^2\left(J''(J''+1)+12\right).
\end{equation}
Squaring Eq.(\ref{defJ'}) and averaging the result with the dipole projected state $J$, one can calculate the angle between the angular momenta $J_p$ and $J_n$:
\begin{equation}
\cos({\bf J}_p,{\bf J}_n)^{(1)}_{J}=\frac{\widetilde{J}^{(1)}_{pn;J}(\widetilde{J}^{(1)}_{pn;J}+1)-\widetilde{J}^{(1)}_{p;J}(\widetilde{J}^{(1)}_{p;J}+1)-\widetilde{J}^{(1)}_{n;J}
(\widetilde{J}^{(1)}_{n;J}+1)}{2\sqrt{\widetilde{J}^{(1)}_
{p;J}(\widetilde{J}^{(1)}_{p;J}+1)\widetilde{J}^{(1)}_{n;J}(\widetilde{J}^{(1)}_{n;J}+1)}}.
\end{equation}
\renewcommand{\theequation}{4.\arabic{equation}}
\setcounter{equation}{0}
\section{A possible extension of the GCSM}
Here we shall consider a particle-core interacting system described by the following Hamiltonian:
\begin{eqnarray}
H&=&H_{GCSM}+\sum_{\alpha}\epsilon_{a}c^{\dag}_{\alpha}c_{\alpha}-\frac{G}{4}P^{\dag}P
\nonumber\\
&-&\sum_{\tau =p,n}X^{(\tau)}_{pc}\sum_{m}q_{2m}\left(b^{\dag}_{\tau,-m}+(-)^mb_{\tau m}\right)(-)^m -X_{sS}{\bf J}_F\cdot{\bf J}_c, 
\label{modelH}
\end{eqnarray}
with the notation for the particle quadrupole operator:
\begin{eqnarray}
q_{2m}&=&\sum_{a,b}Q_{a,b}\left(c^{\dag}_{j_a}c_{j_b}\right)_{2m},\nonumber\\
Q_{a,b}&=&\frac{\hat{j}_{a}}{\hat{2}}\langle j_{a}||r^2Y_2||j_b\rangle .
\end{eqnarray}
Here $H_{GCSM}$ denotes the phenomenological Hamiltonian described in previous section, associated to a proton and neutron bosonic core. The next two terms stand for a set of particles moving in a spherical shell model mean-field and interacting among themselves through pairing interaction. 
The low indices $\alpha$ denote the set of quantum numbers labeling the spherical single particle shell model states, i.e. $|\alpha\rangle =|nljm\rangle =|a,m\rangle$.
The last two terms denoted hereafter as $H_{pc}$ express the interaction between the satellite particles and the core through a quadrupole-quadrupole and a spin-spin force, respectively. The angular momenta carried by the core and particles are denoted by ${\bf J}_c (= {\bf J}_{pn}$) and ${\bf J}_F$, respectively. 
The mean field plus the pairing term is quasi-diagonalized by means of the Bogoliubov-Valatin transformation:
\begin{eqnarray}
a^{\dag}_{\alpha}&=&U_ac^{\dag}_{\alpha}-V_as_{\alpha}c_{-\alpha},\;\;s_{\alpha}=(-)^{j_{\alpha}-
m_{\alpha}},\nonumber\\
a_{\alpha}&=&U_ac_{\alpha}-V_as_{\alpha}c^{\dag}_{-\alpha},\;\;(-\alpha)=(a,-m_{\alpha}).
\end{eqnarray}
The free quasiparticle term is $\sum_{\alpha}E_{a}a^{\dag}_{\alpha}a_{\alpha}$, while the qQ interaction preserves  the above mentioned form, with the factor $q_{2m}$ changed to:
\begin{eqnarray}
q_{2m}&=& \eta^{(-)}_{ab}\left(a^{\dag}_{j_a}a_{j_b}\right)_{2m}+\xi^{(+)}_{ab}\left((a^{\dag}_{j_a}a^{\dag}_{j_b})_{2m}-(a_{j_a}a_{j_b})_{2m}\right),\;\; \rm{where}
\nonumber\\
\eta^{(-)}_{ab}&=&\frac{1}{2}Q_{ab}\left(U_aU_b-V_aV_b\right),\;\;
\xi^{(+)}_{ab}=\frac{1}{2}Q_{ab}\left(U_aV_b+V_aU_b\right).
\end{eqnarray}
We restrict the single particle space to a single-j state where two particles are placed.
In the space of the particle-core states we, therefore, consider the basis defined by:
\begin{eqnarray}
&&|BCS\rangle\otimes\varphi^{(1)}_{JM},\nonumber\\
\Psi^{(2qp;J1)}_{JI;M}&=&N^{(2qp;J1)}_{JI}\sum_{J'}C^{J\;J'\;I}_{J\;1\;J+1}\left(N^{(1)}_{J'}\right)^{-1}\left[(a^{\dag}_ja^{\dag}_j)_J|BCS\rangle\otimes\phi^{(1)}_{J'}\right]_{IM},
\label{basis}
\end{eqnarray}
where $|BCS\rangle$ denotes the quasiparticle vacuum, while $N_{JI}$ is the norm given by
\begin{equation}
\left(N^{(2qp;J1)}_{JI}\right)^{-2}=\sum_{J'}2\left(N^{(1)}_{J'}\right)^{-2}\left(C^{J\;J'\;I}_{J\;1\;J+1}\right)^2.
\end{equation}
The matrix elements of the model Hamiltonian H are given analytically in Appendix A.

Now let us analyze the  proton and neutron angular momentum composition for the two quasiparticle components of the particle-core basis.
The effective angular momenta can be easily calculated:
\begin{eqnarray}
&&\widetilde{J}^{(1)}_{\tau;JI}(\widetilde{J}^{(1)}_{\tau;JI}+1)=\langle \Psi^{(2qp;J1)}_{JI}|\hat{J}^{2}_{\tau}|\Psi^{(2qp;J1)}_{JI}\rangle 
\nonumber\\
&=&\left(N^{(2qp;J1)}_{JI}\right)^2\sum_{J'}2\left(C^{J\; J'\; I}_{J\; 1\; J+1}\right)^2\left(N^{(1)}_{J'}\right)^{-2}
\widetilde{J}_{\tau;J'}(\widetilde{J}_{\tau;J'}+1),\;\tau=p,n ,\nonumber\\
&&\widetilde{J}^{(1)}_{pn;JI}(\widetilde{J}^{(1)}_{(pn;JI}+1)=\langle \Psi^{(2qp;J1)}_{JI}|(\hat{J}_{p}+
\hat{J}_n)^2|\Psi^{(2qp;J1)}_{JI}\rangle \nonumber\\
&=&\left(N^{(2qp;J1)}_{JI}\right)^2\sum_{J'}2\left(C^{J\; J'\; I}_{J\; 1\; J+1}\right)^2\left(N^{(1)}_{J'}\right)^{-2}
\widetilde{J}^{(1)}_{pn;J'}(\widetilde{J}^{(1)}_{pn;J'}+1).
\end{eqnarray}
The angle between proton and neutron angular momenta can be obtained from the equation:
\begin{equation}
\cos({\bf J}_p,{\bf J}_n)^{(1)}_{JI}=\frac{\widetilde{J}^{(1)}_{pn;JI}(\widetilde{J}^{(1)}_{pn;JI}+1)-\widetilde{J}^{(1)}_{p;JI}(\widetilde{J}^{(1)}_{p;JI}+1)-\widetilde{J}^{(1)}_{n;JI}(\widetilde{J}^{(1)}_{n;JI}+1)}{2\sqrt{\widetilde{J}^{(1)}_{p;JI}(\widetilde{J}^{(1)}_{p;JI}+1)\widetilde{J}^{(1)}_{n;JI}(\widetilde{J}^{(1)}_{n;JI}+1)}}.
\end{equation}
\renewcommand{\theequation}{5.\arabic{equation}}
\setcounter{equation}{0}
\section{About the chiral symmetry}
\begin{figure}[h!]
\begin{center}
\vspace*{-1cm}
\includegraphics[width=0.6\textwidth]{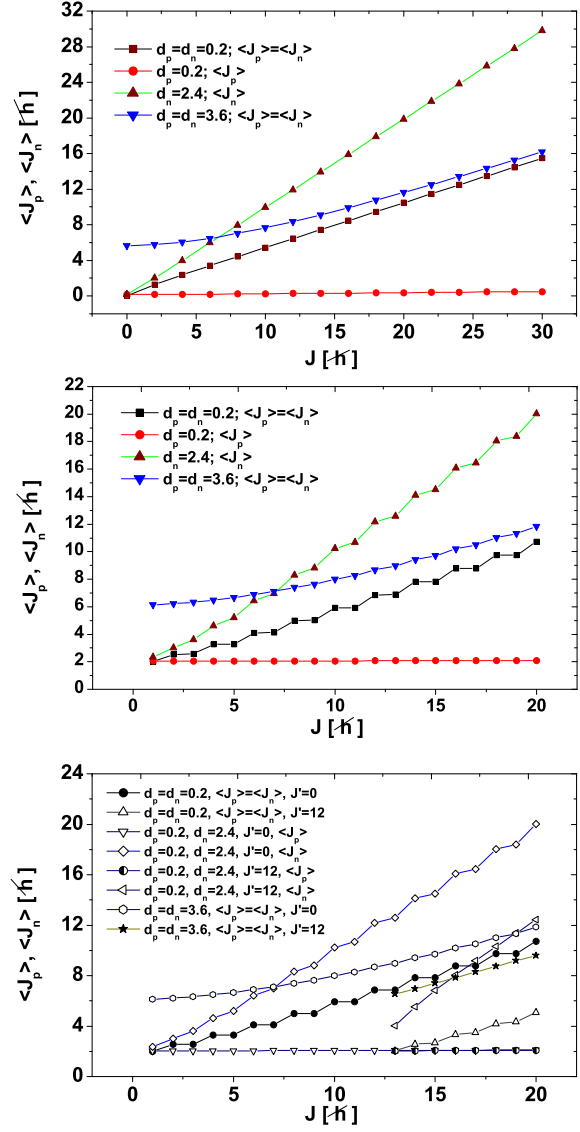}
\end{center}
\vspace*{-1.5cm}
\caption{\scriptsize{Proton and neutron angular momentum composition of the states from the ground band (upper panel), the pure phenomenological dipole band (middle panel) and the two quasiparticle-dipole band (bottom panel). The curves with the symbols of full circles and triangle up respectively, in the upper and middle panels, correspond to $d_p=0.2$ and $d_n=2.4$.} }    
\end{figure}

\begin{figure}[t!]
\begin{center}
\includegraphics[width=0.5\textwidth]{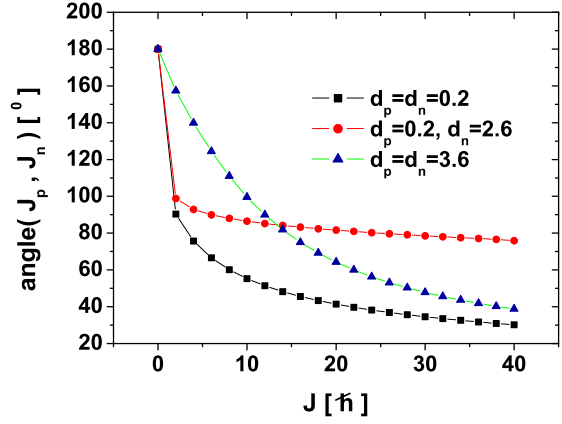}
\end{center}
\vspace*{-1.5cm}
\caption{The angle between ${\bf J}_p$ and ${\bf J}_n$ within the ground-band states $\phi^{(g)}_{JM}$ for three sets of deformations $(d_p, d_n)$. }
\end{figure}
\begin{figure}[h!]
\vspace*{-2cm}
\begin{center}
\includegraphics[width=0.5\textwidth]{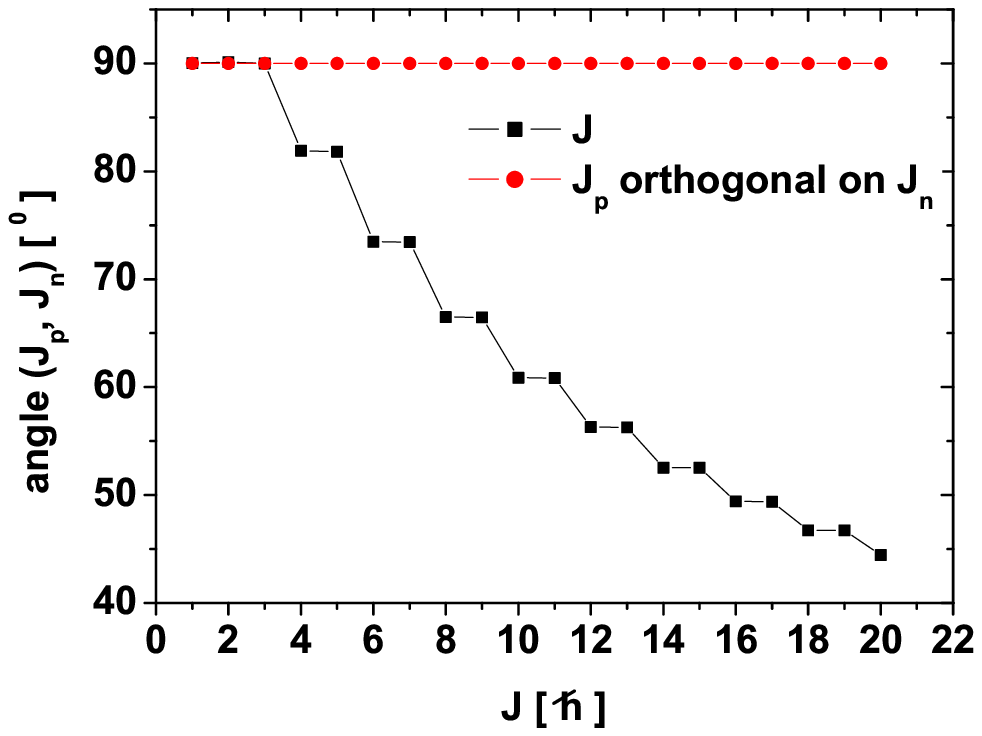}
\end{center}
\vspace*{-1.5cm}
\caption{The angle between ${\bf J}_p$ and ${\bf J}_n$ within the boson dipole state $\phi^{(1)}_{JM}$. The deformation parameter $d$ (see Eq. 2.3) is  equal to 0.2.}    
\end{figure}
\begin{figure}[h!]
\vspace{-2cm}
\begin{center}
\includegraphics[width=0.5\textwidth]{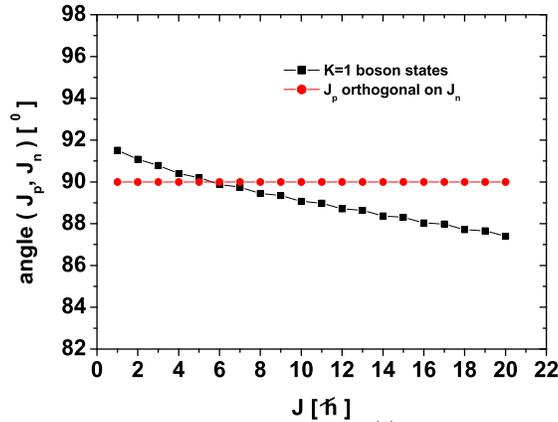}
\end{center}
\vspace*{-1.5cm}
\caption{\scriptsize{The angle between ${\bf J}_p$ and ${\bf J}_n$ within the boson dipole state $\phi^{(1)}_{JM}(d_p,d_n)$. The deformation parameters are $d_p=0.2$  and $d_n=2.4$.}}    
\end{figure}
\clearpage
It is worth studying the separate contribution of protons and neutrons to the total angular momentum of a state belonging to the ground band, to the pure phenomenological dipole band and to 
two quasiparticle-dipole band, respectively. For the three bands this was analytically given by Eqs. (3.9), (3.13) and (4.7),  and plotted in upper, middle and bottom panels of Fig.1 respectively.
Therein, the notations $<J_{\tau}>$ stay for $\widetilde{J}^{(g)}_{\tau J}$, $\widetilde{J}^{(1)}_{\tau;J}$ and
$\widetilde{J}^{(1)}_{\tau;JI}$, respectively. Note that for ground band states, when the the proton and the neutron deformations are equal and large, the two angular momenta are aligned to each other in states of high angular momentum. Indeed, as seen from the upper panel for large $J$ we have $J\approx 2\langle J_p\rangle$.
If the two deformations are very different then, by far, the largest contribution is brought by the most deformed system the weakly deformed subsystem bringing an almost vanishing average angular momentum. As for the pure phenomenological dipole band, represented in the middle panel of Fig. 1, we note an even-odd staggering for small and moderate deformation. Such a structure is washed out for large deformation.
These features are met also for the case of two quasiparticle-dipole states when the two quasiparticles  total angular momentum is equal to zero. Due to the large K quantum number of the two quasiparticle components, when the angular momentum carried by the two quasiparticles is equal to 12, the dipole band starts with the angular momentum 13.

The two quasiparticle-dipole state components of the particle-core basis involve three angular momenta, ${\bf J}_p,{\bf J}_n$, and the quasiparticles total angular momentum denoted by
${\bf J}_{F}$, which, in certain states, could be
mutually orthogonal. Under this circumstance, suppose that the vectors set ${\bf J}_p,{\bf J}_n,{\bf J}_{F}$ form a right trihedral. 

{\it The transformation which changes the orientation of one component of the set, i.e. the right trihedral is transformed into a left one, is conventionally called  chiral. Obviously, any such a transformation may be written as a product of a rotation of angle $\pi$ around a chosen trihedral axis and the space reversal transformation. Excepting the spin-spin term, the Hamiltonian introduced in the previous section is invariant to any chiral transformation. In fact, the chiral symmetry breaking mentioned above is generating the so called chiral bands characterized, first of all, by a large intra-band M1 transition probability.}

The goal of this section is to identify states $\Psi^{(2qp;J1)}_{JI;M}$ characterized by an orthogonal trihedral $({\bf J}_p, {\bf J}_n,{\bf J}_F)$.

The angle between the angular momenta carried by protons and neutrons in a ground band projected state is represented as function of the angular momentum $J$ for different sets of proton and neutron deformations, in Fig. 2. Irrespective of the deformations magnitude, for $J=0$, the angular momenta ${\bf J}_p$ and ${\bf J_n}$ are anti-aligned. For $J=2$ the angle jumps down to $90^0$ and
$98^0$ when both deformations are small or one is small while the other one only moderately small, respectively. Increasing the angular momentum, the angle characterizing the system of small deformations is smoothly decreasing, approaching the aligned picture for very large angular momentum. By contrast, when the proton and neutron deformations are very different, the angle is smoothly but slowly decreasing keeping close to $90^0$. In the case of equal and large proton and neutron deformations the angle is continuously decreasing the rotation gradually aligning the two angular momenta, ${\bf J}_p$ and ${\bf J}_n$.

The relative angle of the proton and neutron angular momenta in the pure boson dipole state $\varphi^{(1)}_{JM}$ is presented in Fig.3.
One notices that the angle is $90^{0}$ in the first three dipole states of angular momenta 1,2 and 3. Increasing the total spin, the corresponding angles decrease monotonically. A step structure for the states $J$ and $J+1$ with J-even shows up. We recall that in our previous applications of the GCSM \cite{1}, the unprojected state $\Psi_g$ was considered for equal deformation parameters for the proton and neutron systems. However, since the number of protons and neutrons are different and, moreover, the two kinds of nucleons occupy different shells, it is reasonable to suppose different deformation parameters for protons and neutrons, respectively. The corresponding projected dipole states are denoted by $\Phi^{(1)}_{JM}(d_p,d_n)$. For this situation, the dependence of the $({\bf J}_p,{\bf J}_n)$ angle on the total angular momentum is presented in Fig. 4.
\newpage
\begin{figure}[t!]
\vspace{-2cm}
\begin{center}
\includegraphics[width=0.5\textwidth]{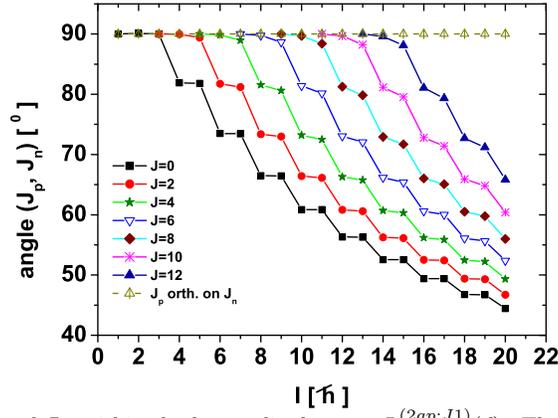}
\end{center}
\vspace*{-1.5cm}
\caption{\scriptsize{The angle between ${\bf J}_p$ and ${\bf J}_n$ within the boson dipole state $\Psi^{(2qp;J1)}_{JI;M}(d)$. The deformation parameter $d$ (see Eq. (2.3)) is equal to $0.2$.}}    
\end{figure}
\begin{figure}[h!]
\vspace{-2cm}
\begin{center}
\includegraphics[width=0.6\textwidth]{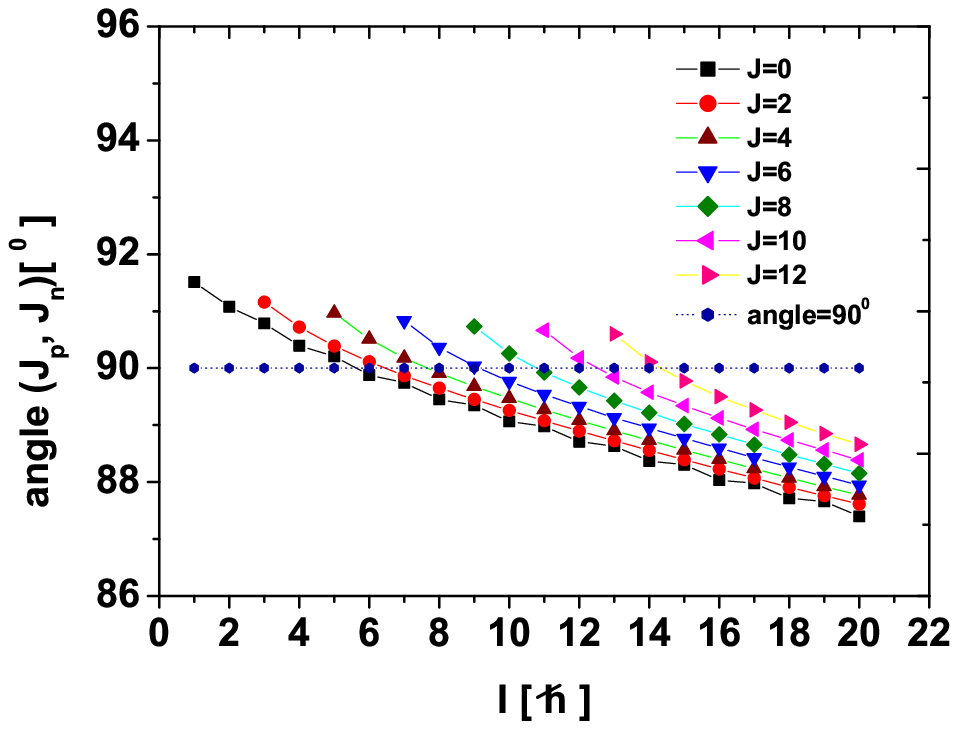}
\end{center}
\vspace*{-1.5cm}
\caption{\scriptsize{The angle between ${\bf J}_p$ and ${\bf J}_n$ within the boson dipole state $\Psi^{(2qp;J1)}_{JI;M}(d_p,d_n)$. The deformation parameters are $d_p=0.2$  and $d_n=2.4$.}}    
\end{figure}
\begin{figure}[h!]
\vspace*{-1cm}
\begin{center}
\includegraphics[width=0.5\textwidth]{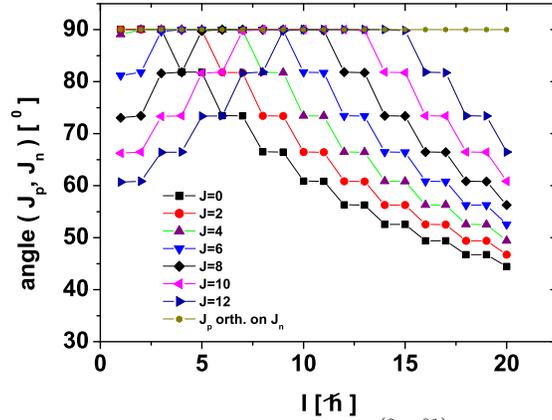}
\end{center}
\vspace{-1.5cm}
\caption{\scriptsize{The angle between ${\bf J}_p$ and ${\bf J}_n$ within the boson dipole state $\Psi^{(2qp;01)}_{JI;M}(d)$. The deformation parameter for protons is equal to that for neutrons and 
$d =0.2$ (see Eq. (2.3)).}}    
\end{figure}
\clearpage
When the deformation for protons is different from that of neutrons, the step structure is washed out and the total angular momenta, where the relative angle is about $90^0$, are shifted to 5, 6 and 7. The angle decreases with angular momentum but with a much lower slope. Indeed, in the considered angular momentum interval the angle varies between $91.5^0$ and $87^0$.

Remarkable the fact that the angle of the proton and neutron angular momenta in the dipole states given in Figs. 3 and 4 is different from that characterizing the ground band states  and shown in Fig. 2 for three sets of the proton and neutron deformation parameters, $(d_p,d_n)$. 
Note that for the state $0^+$, heading the ground band, the two angular momenta, ${\bf J}_p,{\bf J}_n$, are equal in magnitude, have the same direction but different orientation. This property holds irrespective of the deformation parameters $d_p, d_n$. From the value of $180^0$, the angle is decreasing when the total angular momentum is increased. When the proton and neutron deformations are equal the angle tends to zero for J very large. The alignment is reached faster for small deformations than for large ones. If the deformations are different, namely one is small and the other moderately large, the angle is very slowly decreasing for $J\ge 2$, otherwise keeping close to 90$^0$, reflecting the fact that for small deformation the rotational axis is almost indefinite. As for the dipole band, to  build up the dipole state $1^+$ one gets contribution not only  from the ground band state $0^+$, but also from the state $2^+$ which results, for small deformations, an angle between proton and neutron angular momenta close to $90^0$ (see Fig. 3). By contrast, when the deformation is large the above mentioned angle should be between $180^0$ and $160^0$ and, moreover, closer to one or another extreme depending on the rate of the mixture of the states $0^+_g$ and $2^+_g$ in the structure of the dipole state $1^+$. According to this picture, the state $1^+$ is not a typical scissors state, where the angle between the proton and neutron symmetry axes is very small, but rather a shear mode.

Let us see now, how this picture modifies when we add to the boson dipole states the two quasiparticle state factor. As shown in Fig. 5, the case of common small deformation for
protons and neutrons is similar to that from Fig. 3 where the two quasiparticle factor is missing.
By contrast, here we have seven sets of states distinguished by the angular momentum $J$ carried by the quasiparticle component. Otherwise, the step function structure as well as the decreasing behavior as function of the total angular momentum, $I$, are preserved by any of the seven sets. The seven curves differ from each other by the angular moment I, where the protons and neutron angular momenta are orthogonal. Thus, for a given J (=0, 2, 4,..., 12) the total angular momenta for which the proton neutron angle is $90^0$ are I=J+1,J+2,J+3. 
The same remark  holds also for Fig. 6, when compared with the situation from Fig.4. Indeed, it seems that the larger the difference between proton and neutron deformations, the smaller the departure of the $({\bf J}_p,{\bf J}_n)$ angle from $90^0$ and the less pronounced the step structure of the angle I-dependence.

From Fig. 5 it is clear that for each value of the two quasiparticle angular momentum there are three states, the lowest angular momentum states being characterized by an orthogonal 
configuration $({\bf J}_p,{\bf J}_n)$. Since the K quantum numbers for proton and neutron systems included in the core are small and, moreover, the total K being equal to unity, it is reasonable to suppose that ${\bf J}_p$ and ${\bf J}_n$ are both perpendicular to the intrinsic symmetry axis, that is OZ. 
The symmetry axis of the particle motion is determined by the mean field caused by the particle-core interaction of the $qQ$ type. On the other hand, the quasiparticle angular momentum  projection on the symmetry axis is, by construction,  maximal. Therefore, ${\bf J}_F$ 
is oriented along the axis $OZ$, which results in having an orthogonal trihedral $({\bf J}_p,{\bf J}_n,{\bf J}_F)$. Invoking the arguments of Ref.\cite{Frau}, for such states a large transverse dipole moment is expected, which may induce a large M1 transition rate. {\it If one ignores the spin-spin interaction term, the resulting Hamiltonian is invariant to changing the orientation of one of the trihedral component, which means that this Hamiltonian exhibits a chiral symmetry. The spin-spin interaction breaks the chiral symmetry and, therefore, lifts the associated degeneracy. By successively changing the orientation of one trihedral component, one obtains four distinct Hamiltonians and therefore one expects four bands. Each of these bands may be related to the remaining three bands by specific chiral transformation, respectively.} These features are in detail studied in what follows. 

However, before doing that let us consider the states with the quasiparticle factor state with angular momentum  and projection $(J,0)$:
\begin{equation}
\Psi^{(2qp;01)}_{JI;M}={\cal{N}}^{(2qp;01)}_{JI}\sum_{J'}C^{J\; J'\; I}_{0\;\;1\;\;1}\left[(a^{\dagger}_ja^{\dagger}_j)_{J}\varphi^{(1)}_{J'}\right]_{IM}\left(N^{(1)}_{J'}\right)^{-1}.
\end{equation}
In such a state, the three angular momenta, ${\bf J}_p,{\bf J}_n, {\bf J}_F$ are in  the same plane.
Hence, one expects the magnetic properties to be different from those characterizing the state where the mentioned vectors are mutually orthogonal. For comparison, these states are also considered in Figs. 7 and 8. 

\renewcommand{\theequation}{6.\arabic{equation}}
\setcounter{equation}{0}
\section{Magnetic dipole transitions} 
The magnetic moment of the phenomenological core is defined by:
\begin{equation}
{\bf \mu}_c=g_p{\bf J}_p+g_n{\bf J}_n\equiv g_c{\bf J}_{pn},
\end{equation}
where $g_p$, $g_n$ and $g_c$ denote the gyromagnetic factors for proton neutrons and the core.
Multiplying this  with ${\bf J}_c={\bf J}_{pn}$, and averaging the result with the function
$\Psi^{(2qp;J1)}_{JI;M}$, one obtains an equation determining $g_c$:
\begin{equation}
g_{c;JI}=\frac{g_p+g_n}{2}+\frac{g_p-g_n}{2}\frac{\widetilde{J}^{(1)}_{p;JI}(\widetilde{J}^{(1)}_{p;JI}+1)- 
\widetilde{J}^{(1)}_{n;JI}(\widetilde{J}^{(1)}_{n;JI}+1)}{\widetilde{J}^{(1)}_{pn;JI}(\widetilde{J}^{(1)}_{pn;JI}+1)}.
\label{core_g}
\end{equation}
Note that since the deformation parameters for proton and neutron are equal with each other, the average values of proton and neutron angular momenta are the same, $\widetilde{J}^{(1)}_p =\widetilde{J}^{(1)}_n$, which results in having a simple expression for the core gyromagnetic factor:
\begin{equation}
g_c=\frac{g_p+g_n}{2}.
\end{equation}
The expression \ref{core_g} can be easily derived by expressing first the core magnetic moment as a linear combination of the sum and the difference of proton and neutron angular momenta:
\begin{equation}
{\bf \mu}_c = \frac{g_p+g_n}{2}\left({\bf J}_p+{\bf J}_n\right)+\frac{g_p-g_n}{2}\left({\bf J}_p-{\bf J}_n\right).
\end{equation}
Since the scissors state, $1^+$, is antisymmetric with respect to the proton neutron permutation, while the ground state, $0^+$,  is symmetric,  only the second term from the above equation contributes to the transition  $0^+\to 1^+$. This feature is not preserved when we treat the intra transitions of the chiral band, the states participating to  the transition behaving similarly at the proton-neutron permutation.

Denoting by $g_F$ the gyromagnetic factor for the two quasiparticle factor state and following a similar procedure as above we get for the whole system the following gyromagnetic factor:

\begin{equation}
g_{JI}=\frac{g_F+g_c}{2}+\frac{g_c-g_F}{2}\frac{\widetilde{J}^{(1)}_{pn;JI}(\widetilde{J}^{(1)}_{pn;JI}+1)-
J(J+1)}{I(I+1)}.
\end{equation}
We note that both gyromagnetic factors for the core and for the whole system depend on the angular momenta $J$ and $I$.

In order to calculate the M1 transition probability we need the following reduced matrix elements:
\begin{eqnarray}
&&\langle\Psi^{(2qp;J1)}_{JI}||J_F||\Psi^{(2qp;J1)}_{JI'}\rangle=2\hat{I'}\hat{J}\sqrt{J(J+1)}N_{JI}N_{JI'}
\sum_{J_1}\left(N^{(1)}_{J_1}\right)^{-2}\left(C^{J\;J_1\;I}_{J\;1\;J+1}\right)^2W(I'J_11J;JI),\nonumber\\
&&\langle\Psi^{(2qp;J1)}_{JI}||g_pJ_p+g_nJ_n||\Psi^{(2qp;J1)}_{JI'}\rangle =N_{JI}N_{JI'}\hat{I'}\hat{1}\sum_{J_1}C^{J\;J_1\;I}_{J\;1\;J+1}C^{J\;J_1\;I'}_{J\;1\;J+1}\left(N^{(1)}_{J_1}\right)^{-2} W(JJ_1I1;I'J_1)
\nonumber\\
&\times&\left(g_p\sqrt{\widetilde{J}_{p;J_1}(\widetilde{J}_{p;J_1}+1)}+g_n\sqrt{\widetilde{J}_{n;J_1}(\widetilde{J}_{n;J_1}+1)}\right).
\end{eqnarray}
Using the previous results regarding the average value of $\hat{J}^2_{\tau}$, the last expression of the above equations considered for the case $I'=I$,  simplifies to:
\begin{equation}
\langle\Psi^{(2qp;J1)}_{JI}||g_pJ_p+g_nJ_n||\Psi^{(2qp;J1)}_{JI}\rangle =
g_p\sqrt{\widetilde{J}_{p;JI}(\widetilde{J}_{p;JI}+1)}+g_n\sqrt{\widetilde{J}_{n;JI}(\widetilde{J}_{n;JI}+1)}.
\end{equation}

The M1 transition operator is defined by:
\begin{equation}
M_{1,m}=\sqrt{\frac{3}{4\pi}}\mu_{1,m}.
\end{equation}
In Refs.\cite{1,2,3} we pointed out a drawback of the phenomenological descriptions of the magnetic states consisting of that the transition operator does not take care of the Hamiltonian model structure, i.e. is independent of the states participating at transition. Therein, we proposed a possible solution for correcting the mentioned drawback.

Indeed, using the classical expression for the magnetic moment:
\begin{equation}
{\bf \mu}_k=\frac{1}{2c}\int\rho_p({\bf r}\times{\bf v})_kd{bf r},
\end{equation}
with $\rho_p$ and ${\bf v}$ denoting the proton charge density and the velocity of an elementary volume of proton matter having the coordinate ${\bf r}$ and integrating on a liquid drop volume whose surface is expressed in terms of the quadrupole coordinates $\alpha_{\mu}$, one arrives at a quadratic expression in coordinates and their time derivatives. The coordinates and their conjugate momenta are quantized by:
\begin{eqnarray} 
\alpha_{p\mu}&=&\frac{1}{k_p\sqrt{2}}\left(b^{\dagger}_{p\mu}+(-)^{\mu}b_{p,-\mu}\right),
\nonumber\\
\dot{\alpha}_{p\mu}&=&\frac{1}{i\hbar}\left[H,\alpha_{p\mu}\right],   
\end{eqnarray}
where $"\;~\dot{}\;~"$ denotes the time derivative operation.
In this way a simple boson expression for the transition operator was obtained:
\begin{equation}
M_{1\mu}=\sqrt{2}\frac{Mc}{\hbar}R_0\mu_N{\cal F}_{\mu}, R_0=1.2A^{1/3},
\end{equation}
where M denotes the proton mass, $\mu_N$ the nuclear magneton and $c$ the light velocity. The reduced form-factor ${\cal F}_{k_p}$ has the expression:
\begin{eqnarray}
&&q{\cal F}_{\mu}=-\frac{i}{\hbar ck_p^2}\left[(A_1+6A_4)\hat{J}_{p\mu}+\frac{A_3}{5}\hat{J}_{n\mu}\right.\nonumber\\
&&\left.+\frac{\sqrt {10}}{4}(A_2-A_1)\left[(b^{\dag}_nb^{\dag}_p)_{1\mu}+(b^{\dag}_nb_p)_{1\mu}+(b^{\dag}_pb_n)_{1\mu}-
(b_nb_p)_{1\mu}\right]\right. \nonumber\\
&+&\left.\sqrt{2}A_3\left[-\frac{1}{\sqrt{10}}(\Omega^{\dag}_n\hat{J}_{p\mu}+\hat{J}_{p\mu}\Omega_n)-
\Omega^{\dag}_{pn}[-(b^{\dag}_pb_n)_{1\mu}+(b_nb_p)_{1\mu}]+[(b^{\dag}_nb^{\dag}_p)_{1\mu}+(b^{\dag}_nb_p)_{1\mu}]\Omega_{np}\right]\right].\nonumber\\
\end{eqnarray}
where $A_i$'s are the structure coefficients involved in Eq. (2.4).
Here $q$ stands for the momentum transfer when a transition, from an initial state of energy $E_i$ to a final state of energy $E_f$, takes place:
\begin{equation}
q=\frac{E_i-E_f}{\hbar c}.
\end{equation}
From the above equations we note that, even in the second order in bosons, the gyromagnetic factors have components different of the angular momenta $\hat{J}_p$ and $\hat{J}_n$, which are proportional to the proton-neutron dipole operators. Although the present formalism is purely a phenomenological one and therefore the magnetic moments of neutrons are not included, due to the proton neutron coupling terms from the model Hamiltonian, the neutron gyromagnetic factor is not vanishing.  

Actually, restricting the expression for the transition operator to the angular momenta, 
the above equation provides  analytical expressions for the proton and neutron system.  For illustration, in Table I we give the results of our calculations for the reduced magnetic dipole transitions between two adjacent states from a two quasiparticle band, for two sets of the deformation parameters. These are chosen such that to correspond to a near vibrational regime. We recall that a rotational picture is reached for a deformation parameter larger than 3 \cite{Rad1}. We note that for $J\ge 6$, where $J$ denotes the quasiparticle total angular momentum,  and system angular momentum $I$ larger than 10, the transitions might be considered of collective nature. Although we truncated the angular momentum $I$ to 20, from Table I it is conspicuous that the larger is I, the larger is the M1 strength.
\begin{table}[h!]
{\scriptsize
\begin{tabular}{|c|ccccccc|ccccccc|}
\hline
$I\to(I-1)$&\multicolumn{7}{c|}{($d_p,d_n$)=(1.0, 1.0)}&\multicolumn{7}{c|}{($d_p,d_n$)=(0.2, 2.4)}\\
\hline
I         &J=0&2  &  4 &  6 &  8 & 10 & 12 &J=0 &  2 &  4 &  6 &  8 &  10&12\\
\hline 
2 &0.929& & & & & & &0.691 &  &  &  &  &  &  \\
3 &0.720 &       &       &       &       &       &       & 0.535 &       &       &       &       &       &       \\
4 &0.765 & 0.057 &       &       &       &       &       & 0.468 & 0.112 &       &       &       &       &       \\
5 &0.669 & 0.158 &       &       &       &       &       & 0.409 & 0.248 &       &       &       &       &       \\ 
6 &0.773 & 0.216 & 0.169 &       &       &       &       & 0.393 & 0.346 & 0.367 &       &       &       &       \\
7 &0.704 & 0.287 & 0.438 &       &       &       &       & 0.361 & 0.415 & 0.786 &       &       &       &       \\
8 &0.832 & 0.297 & 0.648 & 0.280 &       &       &       & 0.362 & 0.463 & 1.110 & 0.656 &       &       &       \\
9 &0.773 & 0.358 & 0.833 & 0.722 &       &       &       & 0.340 & 0.500 & 1.353 & 1.402 &       &       &       \\
10&0.913 & 0.335 & 0.950 & 1.104 & 0.376 &       &       & 0.350 & 0.524 & 1.538 & 2.011 & 0.939 &       &       \\ 
11&0.858 & 0.400 & 1.073 & 1.437 & 0.979 &       &       & 0.333 & 0.547 & 1.679 & 2.491 & 2.014 &       &       \\
12&1.004 & 0.352 & 1.131 & 1.692 & 1.531 & 0.459 &       & 0.346 & 0.557 & 1.789 & 2.877 & 2.938 & 1.204 &       \\
13&0.951 & 0.427 & 1.224 & 1.921 & 2.023 & 1.206 &       & 0.332 & 0.575 & 1.876 & 3.184 & 3.681 & 2.593 &       \\
14&1.102 & 0.359 & 1.242 & 2.087 & 2.429 & 1.916 & 0.531 & 0.348 & 0.576 & 1.945 & 3.432 & 4.301 & 3.811 & 1.447 \\ 
15&1.050 & 0.446 & 1.322 & 2.250 & 2.787 & 2.565 & 1.404 & 0.335 & 0.593 & 2.000 & 3.635 & 4.814 & 4.845 & 3.130 \\
16&1.204 & 0.359 & 1.313 & 2.356 & 3.078 & 3.124 & 2.259 & 0.352 & 0.585 & 2.044 & 3.802 & 5.242 & 5.721 & 4.641 \\ 
17&1.152 & 0.459 & 1.388 & 2.478 & 3.341 & 3.622 & 3.057 & 0.340 & 0.603 & 2.082 & 3.941 & 5.601 & 6.464 & 5.955 \\
18&1.308 & 0.356 & 1.356 & 2.544 & 3.550 & 4.047 & 3.766 & 0.359 & 0.589 & 2.110 & 4.057 & 5.905 & 7.099 & 7.094 \\
19&1.257 & 0.470 & 1.434 & 2.641 & 3.748 & 4.428 & 4.406 & 0.348 & 0.614 & 2.140 & 4.155 & 6.164 & 7.644 & 8.080 \\
20&1.415 & 0.354 & 1.383 & 2.677 & 3.899 & 4.751 & 4.968 & 0.368 & 0.609 & 2.161 & 4.235 & 6.382 & 8.113 & 8.938 \\
\hline 
\end{tabular}}
\caption{The BM1 values, given in units of $\mu_N^2$, of the transitions $I\to (I-1)$ calculated with the wave functions $\Psi^{(2qp;J1)}_{JI;M}$ given by Eq.(\ref{basis}), for two sets of deformation parameters $(d_p,d_n)$. The magnetic dipole transition operator is determined by the following gyromagnetic factors: $g_F=1.3527 \mu_N$; $g_p=0.666\mu_N$; $g_n=0.133\mu_N$. }
\end{table}

\renewcommand{\theequation}{7.\arabic{equation}}
\setcounter{equation}{0}
\section{Numerical results and discussions}
The formalism described in the previous sections was applied for $^{192}$Pt. Unfortunately there are no available data concerning the magnetic bands for even-even-nuclei. In choosing the nucleus of $^{192}$Pt we had in mind that the Pt isotopes around A=192 are gamma soft nuclei and a phase transition from prolate to oblate through a triaxial shape is expected to occur for $^{192}$Pt.
Indeed, the signature for a triaxial rotor 
\begin{equation}
E_{2^+_g}+E_{2^+_{\gamma}}=E_{3^{+}_{\gamma}},
\end{equation}
is satisfied with a good accuracy by the chosen nucleus. The left hand side of the above equation amounts of 929 keV, which should be compared with the value of the right hand side, which is 921 keV.
As noticed by many authors the triaxial shapes favor the occurrence of chiral configurations.

We calculated first the excitation energies for the bands described by the angular momentum projected functions
$\phi^{(g)}_{JM}|BCS\rangle, \phi^{(\beta)}_{JM}|BCS\rangle, \phi^{(\gamma)}_{JM}|BCS\rangle, \phi^{(1)}_{JM}|BCS\rangle, \widetilde{\phi}^{(1)}_{JM}|BCS\rangle $ (2.1) and 
$\Psi^{(2qp;J1)}_{JI;M}$ (4.5 ) and the particle-core Hamiltonian $H$ (4.1). Several parameters like the structure coefficients defining  the model Hamiltonian and the deformation parameters are to be fixed. Since in the present application the proton and neutron deformations are equal, we need only one "global" deformation, $\rho =\sqrt{2}d$. For a given $\rho$ we determine the parameters involved in $H_{GCSM}$ by fitting the excitation energies in the  ground, $\beta$ and $\gamma$ bands, through a least square procedure. We varied then $\rho$ and kept those value which provides the minimal root mean square of the results deviations from the corresponding experimental data. Excitation energies of the phenomenological magnetic bands described by $\phi^{(1)}_{JM}$ and 
$\widetilde{\phi}^{(1)}_{JM}$ respectively are free of any adjusting parameters. The strengths of the pairing and $Q.Q$ interaction were taken close to the values used in Ref. \cite{Lima}, where spectra of some Pt even-even isotopes  where interpreted  with a particle core Hamiltonian, the core being described by the Coherent State Model (CSM). Thus, the quasiparticle energy is 1.25 MeV while the strength $X'_{pc}$ defined by:
\begin{equation}
X'_{pc}=6.5\eta^{(-)}_{\frac{11}{2}\frac{11}{2}}\frac{\hbar}{M\omega_0}X^{(p)}_{pc},
\label{xprime}
\end{equation}
 is taken to be -0.023 MeV. The notations $M$ and $\omega_0$ are used for nucleon mass and the shell model single particle frequency. Since the considered outer particles are protons, the neutron particle-core coupling term is ineffective. Therefore we put $X^{(n)}_{pc}=0.$
The parameters mentioned above have the values listed in Table II.
\begin{table}[h!]
\begin{tabular}{|ccccccc|}
\hline
$\rho=d\sqrt{2}$~~   & ~~   $A_1$  ~~ &~~     $A_2$ ~~   & ~~   $A_3$ ~~   & ~~   $A_4$ ~~&~~  $X'_{pc}$ ~~ &~~  $X_{sS}$\\
\hline
  2.0   &   555.4   & -25.4   &   -12.8  &   7.7  &  -23.4  &  1. \\
\hline
\end{tabular}
\caption{ The structure coefficients of the model Hamiltonian (4.1) determined as described in the text, are given in units of keV. The deformation parameter $\rho$ is a-dimensional. The parameter $X'_{pc}$ is that defined by Eq. (\ref{xprime}). }
\end{table}
\begin{figure}[t!]
\vspace{-2cm}
\begin{center}
\includegraphics[width=0.5\textwidth]{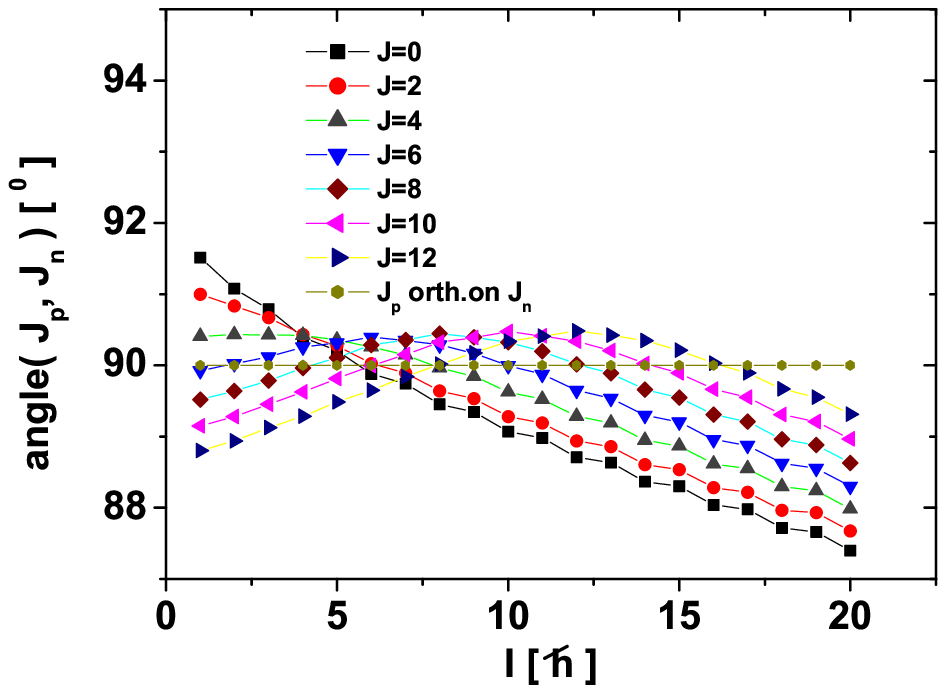}
\end{center}
\vspace{-1.5cm}
\caption{\scriptsize{The angle between ${\bf J}_p$ and ${\bf J}_n$ within the boson dipole state $\Psi^{(2qp;01)}_{JI;M}(d_p,d_n)$. The deformation parameters are $d_p=0.2$  and $d_n=2.4$.}}    
\end{figure}
\begin{figure}[h!]
\vspace*{-2cm}
\begin{center}
\includegraphics[width=0.5\textwidth]{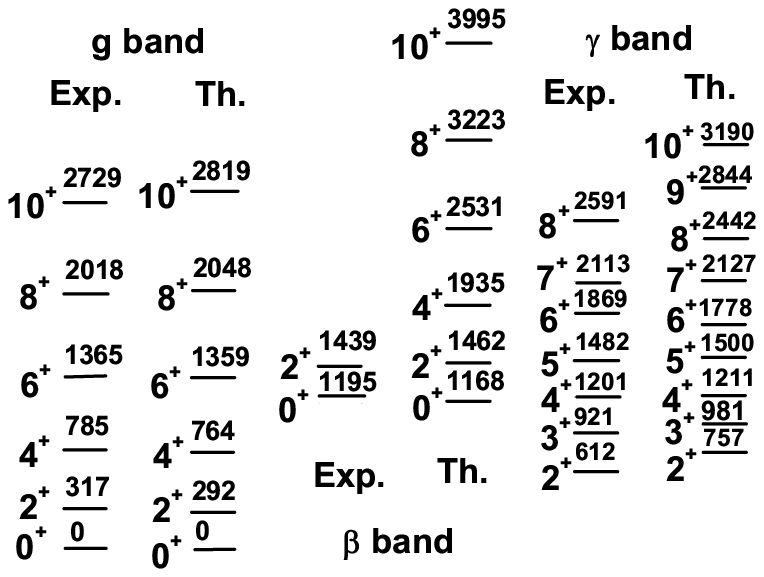}
\end{center}
\vspace*{-1.5cm}
\caption{\scriptsize {Experimental and calculated excitation energies in ground, $\beta$ and $\gamma$ bands for $^{192}$Pt. They correspond to the fitted parameters listed in Table 2. The r.m.s. value of the deviation of the theoretical results and the corresponding experimental data is equal to 67 keV.}}    
\end{figure}
\begin{figure}[h!]
\vspace*{-2cm}
\begin{center}
\includegraphics[width=0.5\textwidth]{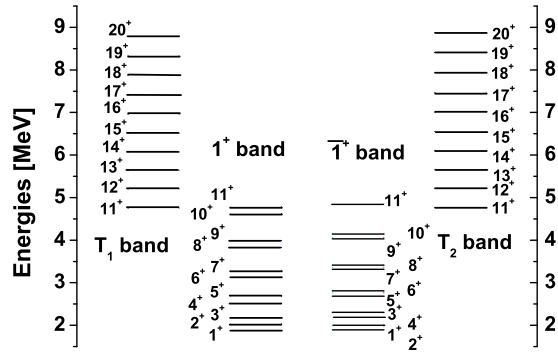}
\end{center}
\vspace*{-1.cm}
\caption{\scriptsize{The excitation energies for the dipole bands described by $\phi^{(1)}_{JM}$ (left lower column) and $\tilde{\phi}^{(1)}_{JM}$ (right lower column), respectively. The bands $T_1$ (upper left column) and $T_2$ (upper right column), conventionally called twin bands,  are also shown. The $T_{1}$ and $T_{2}$ bands were obtained with $X^{\prime}_{pc}$=-0.023 MeV and $X_{sS}$= 0.001 MeV for the left column and $X_{sS}$=-0.001MeV for the right column.}}    
\end{figure}
\clearpage

\begin{figure}[t!]
\vspace*{-1cm}
\begin{center}
\includegraphics[width=0.5\textwidth]{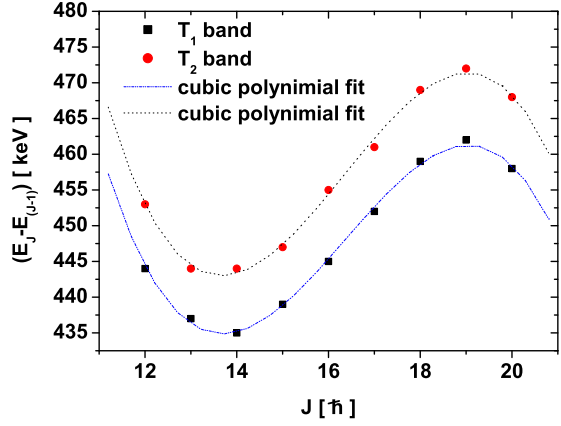}
\end{center}
\vspace*{-1.5cm}
\caption{\scriptsize{Energy spacings in the two twin bands $T_1$ and $T_2$. }}    
\end{figure}
\begin{figure}[h!]
\vspace*{-1.cm}
\begin{center}
\includegraphics[width=0.5\textwidth]{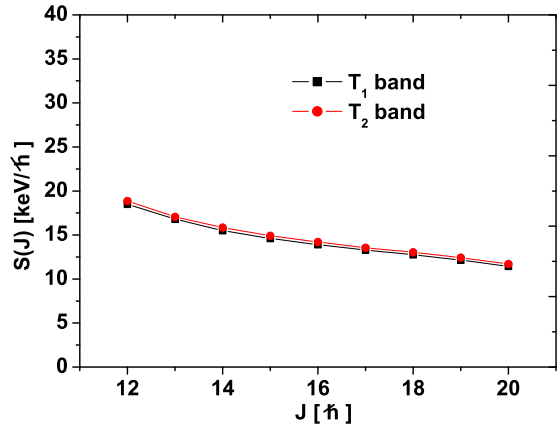}
\end{center}
\vspace{-1.cm}
\caption{\scriptsize{ The signature energy staggering $S(J)$, defined by Eq. (\ref{signat}), is represented as function of the angular momentum J, in the bands $T_1$ and $T_2$.}}    
\end{figure}
\begin{figure}[h!]
\vspace*{-0.5cm}
\begin{center}
\includegraphics[width=0.5\textwidth]{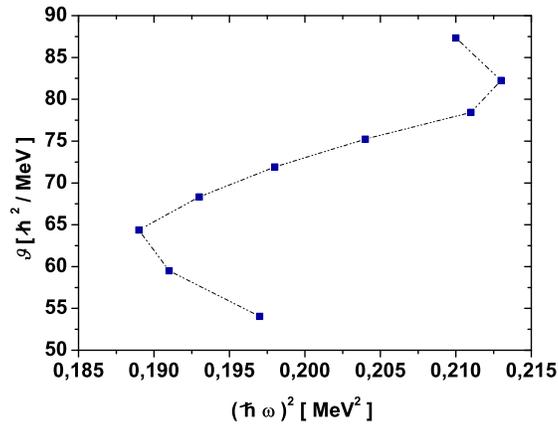}
\end{center}
\vspace*{-1cm}
\caption{\scriptsize{The double moment of inertia calculated for the angular momenta $12^+-20^+$ with Eq.(\ref{inert}) are represented as function of the corresponding rotational frequency given by 
(\ref{freq}).}}    
\end{figure}
\clearpage
Excitation energies calculated with these parameters are compared with the corresponding experimental data, in Fig. 9. One notes a reasonable agreement of results with the corresponding experimental data. The weak feature of our formalism is that does not reproduce the right staggering in the $\gamma$ band. Actually, the experimental energy spacings in this band is almost constant up to the state $5^+$, increases for $6^+$ and then a smaller spacing for the pair of states $6^+, 7^+$ is recorded. Since one has only one staggering situation, one cannot conclude upon a staggering $(J^+, (J+1)^+)$ with $J$-even. It may happen that the state $6^+$ does not really belong to the $\gamma$ band. Thus, to draw a definite conclusion one needs data for excitation energies of the higher spin states. On the other hand the GCSM formalism \cite{1} predicts for small deformation a staggering $(3^+, 4^+); (5^+,6^+); (7^+, 8^+)$, etc. while for large deformation the doublet structure is changed to $(2^+,3^+); (4^+,5^+); (6^+,7^+)$, etc.
The results shown in Fig. 9 is compatible with the first level clustering, which reflects the regime of a small deformation. Indeed, the energy spacings, given in keV, are:
224; 230; 289; 278; 349; 315; 402; 346. As seen in the list, except for the spacing $(3^+,4^+)$ which is almost the same as $(2^+,3^+)$, the rule for the doublet structure $(J^+,(J+1)^+)$ with
$J$ odd is obeyed.

\begin{figure}[h!]
\begin{center}
\includegraphics[width=0.5\textwidth]{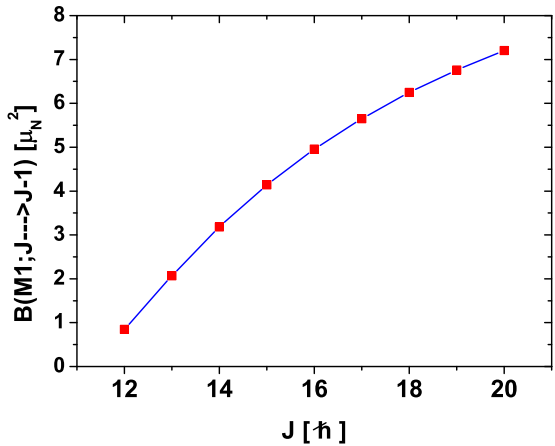}
\end{center}
\vspace*{-1cm}
\caption{\scriptsize{The $B(M1)$  values associated with the dipole magnetic transitions between two consecutive energy levels, in the $T_1$ band. The gyromagnetic factors employed in our calculations are:
$\mu_p=0.666\mu_N$,\;$\mu_n=0.133\mu_N$ and $\mu_F=1.289\mu_N$. As usual the spin gyromagnetic factor was quenched by a factor 0.75 in order to account for the influence of the proton excited states on the magnetic moment.}}    
\end{figure}
%\clearpage

Results for the magnetic dipole bands are plotted in Fig. 10. Excitation energies shown there are those from Table III.
\begin{table}
\begin{tabular}{|c|cccc|}
\hline
$J^+$  ~ & ~    $1^+$-band ~ &~  $\bar{1}^+$-band ~ & ~   $T_1$-band ~ &~   $T_2$-band \\
\hline
$1^+$   &   1.874         &  2.010           &                  & \\
$2^+$   &   2.033         &  1.983           &                  &  \\
$3^+$   &   2.183         &  2.291           &                  &\\
$4^+$   &   2.519         &  2.289           &                  &\\
$5^+$   &   2.676         &  2.763           &                  &  \\
$6^+$   &   3.127         &  2.783           &                  &\\
$7^+$   &   3.287         &  3.364           &                  &\\
$8^+$   &   3.832         &  3.413           &                  &\\
$9^+$   &   3.994         &  4.065           &                  & \\
$10^+$   &  4.623         &  4.147           &                  & \\
$11^+$   &  4.785         &  4.852           &  4.757           &  4.765 \\
$12^+$   &  5.492         &  4.969           &  5.201           &  5.218\\
$13^+$   &  5.651         &  5.718           &  5.638           &  5.662\\
$14^+$   &  6.436         &  5.868           &  6.073           &  6.106\\
$15^+$   &  6.589         &  6.655           &  6.512           &  6.553\\
$16^+$   &  7.450         &  6.840           &  6.957           &  7.008\\
$17^+$   &  7.596         &  7.661           &  7.409           &  7.469\\
$18^+$   &  8.535         &  7.881           &  7.868           &  7.938\\
$19^+$   &  8.670         &  8.735           &  8.330           &  8.410\\
$20^+$   &  9.689         &  8.989           &  8.788           &  8.878 \\
\hline
\end{tabular}
\caption{Excitation energies, given in MeV, for the four magnetic bands denoted by $1^+$, $\bar{1}^+$, $T_1$ and $T_2$, respectively. The twin bands $T_1$ and $T_2$ have $K=11$.}
\end{table}
The lower bands exhibit a pronounced doublet structure. Indeed, in the band $1^+$ we notice the staggering $4^+,5^+; 6^+,7^+; 8^+,9^+; etc.$, while in the band $\bar{1}^+$ the states are grouped in a different manner: $1^+,2^+;3^+,4^+;5^+,6^+;7^+,8^+;etc.$. The first three states of the $1^+$ band are close in energy, while in the band $\bar{1}^+$ the first two doublets have an unnatural  spin ordering. The experimental data \cite{Coral} show two states of uncertain spin assignment which decay by M1 to $2^+_g$, $2^+_{\gamma}$ and $0^+_g$ and lie close to the band heads of the two dipole bands having the energies of 1.881 MeV and 2.048 MeV respectively. According to our calculations these states might have the spin 1 and 2 respectively, the mentioned energies being comparable with those associated to the first two states of the band $T_1$. The lowest dipole states of
magnetic nature are identified as having the energies 2.149 MeV and 2.319 MeV respectively, which are not too far from the calculated energies of the states $1^+$. In order to decide to which of the two experimental sets of data could be associated the results of our calculations, additional investigations are necessary from both theoretical and experimental sides.

In the upper part of Fig. 10 we give the excitation energies of the bands $T_1$ and $T_2$, which are tentatively called twin bands. They have  some specific properties. First of all, both are $K=11$ bands.  The meaning of this statement is as follows. Since the unprojected state, generating the bands $T_1$ and $T_2$ through angular momentum projection, is a $K=11$, after projection the wave function is a superposition of different K components among which the one having $K=11$ prevails over the others \cite{Rad1}. The energies of states of the same angular momentum are close to each other. Indeed, their difference ranges from 8 to 90 keV. It is worth noting that energy spacing varies very little in the two twin bands.  Indeed, in $T_1$ it goes from 435 keV  reached  for $13^+$, to 462 keV met at $19^+$. As for the $T_2$ band the minimum energy spacing is of 444 keV met for three states,
$12^+, 13^+, 14^+$, while the maximum spacing is 472 keV for $19^+$. These spacings were plotted in Fig. 11 as function of angular momentum. The curves for the two twin bands are almost parallel to each other and behave as a polynomial in J, of rank three. These spacings are used to calculate the so called {\it signature energy staggering}, defined by:
\begin{equation}
S(J)=\frac{E(J)-E(J-1)}{2J}.
\label{signat}
\end{equation}
This function, plotted in Fig.12, exhibits no staggering and is decreasing monotonically and  very slowly with J. Indeed, the  e-cart of maximum and minimum value is only of about 7 keV. For an 
ideal chiral band this parameter should be independent of $J$.
Both twin bands intersect the lower dipole bands at the energy level $11^+$. Due to this feature we would expect that a backbending takes place at this angular momentum. However, due to the doublet structure in the lower dipole bands it is difficult to define consistently the moment of inertia for the $\Delta J=1$ states. Despite the mentioned encountered difficulties, the plot of the moment of inertia vs. the rotational frequency squared starts with a backbending, continues, from  $14^+$,  with a forward bending and again a backbending from $19^+$. This picture is common for both twin bands. For illustration, in Fig. 13 we present the situation of the $T_1$ band. Denoting by ${\cal J}$, $E(J)$ and $\omega$ the double  moment of inertia, the energy of the state $J^+$ belonging to the $T_1$ band and the rotational frequency respectively, for the chosen $\Delta J=1$ band one gets:
\begin{eqnarray}
\cal{J}&=&\frac{2(J+1)}{E(J+1)-E(J)},
\label{inert}\\
\hbar\omega &=& E(J+1)-E(J).   
\label{freq}  
\end{eqnarray}
From Fig. 13 we see that, indeed, the moment of inertia exhibits a double backbending when is represented as function of the rotational frequency squared. If we consider also the energy levels of the band $1^+$ before its crossing with the band $T_1$ the graph of Fig. 13 would be continued to the left by a saw teeth like curve.

Finally, the M1 transition probabilities have been calculated with the equations (6.6)-(6.8). The gyromagnetic factors for the collective core, denoted by $g_p$ and $g_n$, were determined
from equations predicted by the GCSM model,
\begin{equation}
g_c=\frac{g_p+g_n}{2},\;\;g_n=\frac{1}{5}g_p,
\end{equation}
and taking $g_c=\frac{Z}{A}$.
The results are represented in Fig.14 as a function of $J$. The $J$ dependence seems to be quadratical, the $B(M1)$ value increasing from 0.847 $\mu_{N}^2$ to 7.204 $\mu_{N}^2$.
We remark that the states used for the description of the excitation energies exhibit a moderate deformation, $\rho=2$. We recall that the application of the GCSM to a wide region of nuclei suggests that the well deformed nuclei are characterized by $\rho \ge 3$. This implies that the large M1 transition probabilities for the states of the twin bands are not caused by a large nuclear deformation as happens in the case of scissors mode, but by the specific angular momenta geometry of the chiral bands. Note that in the present calculations we considered the term of the model Hamiltonian breaking the chiral symmetry only for energies but not for the corresponding wave functions. This feature leads to the fact that the two partner bands are described by identical functions which results in having the same B(M1) values for both.  

Note that the bands $T_1$ and $T_2$ correspond to two reference frames of the three angular momenta ${\bf J}_F, {\bf J}_p, {\bf J}_n$ which are related by a chiral transformation which changes the sign of ${\bf J}_F$. The matrix elements of the $X_{sS}$ term in the two reference frames differ from each other by sign. Therefore, for one band, $T_1$, the interaction $sS$  is attractive while for the other band, $T_2$, repulsive. However, there are another two chiral transformations which change the signs of ${\bf J}_n$ and ${\bf J}_p$, respectively, of the right handed frame associated to the band $T_1$, $F_1$. Each of the corresponding bands is therefore a partner band for $T_1$.  The additional bands will be denoted hereafter by $T_3$ and $T_4$ respectively.
They are also partner bands for $T_2$ since their frames, $F_3$ and $F_4$ are obtainable from that defining $T_2$, $F_2$, by simple transformations. Indeed, $F_3$ can be obtained from $F_2$ by a rotation of angle $\pi$ around ${\bf J}_p$, while $F_4$ is obtainable from $F_2$ by a rotation of an angle equal to $\pi$, around ${\bf J}_n$. $T_3$ and $T_4$ are themselves partner to each other, the associated frames being related by a $\pi$-rotation.  Indeed, $F_4$ is obtainable from $F_3$ by rotating it with the angle $\pi$ around ${\bf J}_F$. However, the $sS$ interaction is not invariant to the mentioned rotations which results that the bands $T_2$, $T_3$ and $T_4$ are different from each other. Note that each of the $\pi$ rotations, mentioned above, is a product of two chiral transformations and therefore a chiral transformation, given the fact that chiral transformations form a group.

We mention again that so far the chiral symmetry has been studied for odd-odd and odd-even nuclei around A=130 \cite{Petrache96,Simon2005} and A=100\cite{Vaman}. Only recently the investigation was extended to some heavy nuclei with $A\approx 190$ \cite{Balab}. Although the first interpretation of the twin bands in terms of a spontaneous chiral symmetry breaking was given by Frauendorf
\cite{Frau97}, the first measurement was performed already one year earlier \cite{Petrache96}. Several approaches devoted to the chiral bands description have been proposed. Among these, the particle-asymmetric rotor (PAR)  model is the most popular. It is interesting to mention that PAR was developed, both analytically and numerically, by one of the authors (A.F., in collaboration), and, moreover, the nuclei studied belong to the regions mentioned above\cite{Fas1,Fas2,Fas3,Fas4}. The experimental systematics established the criteria upon which one could decide whether a pair of bands might be considered of a chiral nature. Briefly, these are: 1) The partner bands are almost degenerate. 2) The energy staggering parameter must be  angular momentum independent. 3) The staggering behavior of the ratio $B(M1)/B(E2)$ and $B(M1)_{in}/B(M1)_{out}$, where  $B(M1)_{in}$ and $B(M1)_{out}$ denote the intra-band and inter-band reduced M1 transition probabilities for the partner bands. In Ref.\cite{Petrache06} it was shown that these criteria are necessary but sometimes not sufficient, the partner bands corresponding to nuclear shapes which are not close to each other.

Note that our procedure is based on angular momentum projection from  proton-neutron boson states. Until now this has been overlooked, since boson Hamiltonians invariant to the rotation transformation were treated with   basis  states of good angular momentum. Due to this feature people focused on angular momentum projection from a many body deformed state ( see for example 
Refs. \cite{Robledo,Egido1,Guzman,Egido2}). Our procedure has the advantage, over the other boson formalisms,  of not having redundant components caused by using a different sets of Euler angles for protons and neutrons, respectively.

To our knowledge the present paper is the first one devoted to the description of the chiral bands in even-even nuclei.

\renewcommand{\theequation}{7.\arabic{equation}}
\setcounter{equation}{0}
\section{Conclusion}

In the previous Sections we formulated a semi-phenomenological model to describe the magnetic bands for even-even nuclei which are almost spherical or moderately deformed. 

The main steps performed towards achieving the goal of this paper can be summarized as follows.

The phenomenological Hamiltonian specific to the GCSM model, previously used to describe the magnetic scissors-like states,  is amended with a particle-core quadrupole-quadrupole and a spin-spin interaction term. The pure single particle term describes a set of nucleons moving in a spherical shell model mean-field and interacting among themselves with pairing force.
The particle-core space is generated by a set of particle-core product functions. The first subset has the GCSM model functions for the ground, $\beta$, $\gamma$, $1^+$ and $\bar{1}^+$ bands as core components, while the particle factor function is the quasiparticle vacuum state denoted by $|BCS\rangle$. The second subset of the particle-core basis consists of a quasiparticle component which is a state of two quasiparticles from the shell $h_{11/2}$ of total angular momentum $J$, with $J=0,2,4,...,10$, and a core component which might be any state of the magnetic band $1^+$.
Angular momentum composition of the projected states suggests that the two quasiparticle-core states may favor a chiral configuration for the angular momenta carried by the three subsystems, i.e., ${\bf J}_{F},{\bf J}_{p}, {\bf J}_{n}$. Moreover, the reduced M1 intra-band transition probabilities acquire large values, although the nuclear deformation places the nuclear system in the region  either  of near vibrational or of a transitional region.

Energies are defined by averaging the model Hamiltonian with the basis states. 
The model Hamiltonian involves a term which breaks the chiral symmetry. Due to this term there are four bands which are related by  specific chiral transformations. Energies for these bands are defined as average values of the model Hamiltonian and its chirally transformed ones with a dipole two quasiparticles coupled to a phenomenological boson dipole band. We note that the chiral bands cross the phenomenological boson dipole band and therefore we expect that several backbending-s will show up.

The parameters involved in the model Hamiltonian  were fixed by fitting the experimental energies in the ground, 
$\beta$ and $\gamma$ bands. The application was made for $^{192}$Pt, the choice being justified by its triaxial features which might favor a chiral geometry for the already mentioned three angular momenta.

The bands denoted by $T_1$ and $T_2$ respectively, exhibit a set of properties which certify their quality of partner bands of a chiral nature: 1) The two bands are almost degenerate; 2) The moment of inertia considered as a function of the rotational frequency squared presents two backbending-s; 3) The signature energy staggering is almost angular momentum independent; 4) The B(M1) values associated to the intra-band transitions are large despite the fact that the deformation is typical for a transitional spherical-deformed region.

Concluding, the present paper proposes a formalism to quantitatively describe the properties of the chiral magnetic bands in even-even nuclei. This was positively tested by the application to the case of $^{192}$Pt.

Our work proves that the mechanism for chiral symmetry breaking which also favor a large transversal component for the dipole magnetic transition operator is not unique.
As a matter of fact there are arguments recommending the mixed systems of quadrupole and octupole bosons and a set of valence nucleons as a good candidate for achieving a chiral configuration
\cite{Rad73,Rad74,RadRadRad}. Such a solution will be in detail studied in a subsequent paper.   

Our description is different from the ones from literature in the following respects. While the previous formalisms deal with odd-odd nuclei here we treat even-even nuclei.
While until now there were only two magnetic bands related by a chiral transformation, here we found four magnetic bands having this property. Here we considered two proton quasiparticle bands but alternatively we could chose two neutron quasiparticles  and one proton plus one neutron quasiparticle bands. Of course, the last mentioned bands would describe an odd-odd system. We already checked that a two neutron quasiparticle band is  characterized by a non-collective M1 transition rate. This feature suggests that, indeed, the orbital magnetic moment carried by protons play an important role in determining a chiral magnetic band. The core is described by angular momentum projected states from a proton and a neutron coherent state as well as from its lowest order polynomial 
excitations. Among the three chiral angular momentum components two are associated to the core and one to a two quasiparticle system. By contradistinction the previous descriptions, devoted to odd-odd systems, use a different picture. The core carries one angular momentum and moreover its shape structure determines the orientation of the other two angular momenta associated to the odd proton and odd neutron, respectively.

Experimental data for chiral bands in even-even nuclei are desirable. These would encourage us to extend the present description to a systematic study of the chiral features in even-even nuclei.

{\bf Acknowledgment.} This work was supported by the Romanian Ministry for Education Research Youth and Sport through the CNCSIS project ID-2/5.10.2011.

\renewcommand{\theequation}{A.\arabic{equation}}
\setcounter{equation}{0}
\section{Appendix A} 

Here we give the analytical expression of the model Hamiltonian matrix elements, corresponding to the basis states \ref{basis}:
\begin{eqnarray}
&&\langle \Psi^{(2qp;J1)}_{JI}|H|\Psi^{(2qp;J1)}_{J_1I}\rangle=-4\hat{2}\hat{J}\hat{J}_1X^{(\tau)}_{pc}N^{(2qp;J1)}_{JI}N^{(2qp;J_11)}_{J_1I}\eta^{(-)}_{jj}W(JjJ_1j;j2)\nonumber\\
&\times&\sum_{J'J''}\hat{J}'C^{J\;J'I}_{J\;1\;J+1}C^{J_1\;J''\;I}_{J_1\;1\;J_1+1}
W(J_1 2 I J'; J J'')\langle \phi^{(1)}_{J'}||b^{\dag}_{\tau}+b_{\tau}||\phi^{(1)}_{J''}\rangle\nonumber\\
&-&X_{sS}\delta_{J,J_1}\left[I(I+1)-J(J+1)-\left(N^{(2qp;J1)}_{JI}\right)^2\sum_{J'}2J'(J'+1)\left(C^{J\;J'\;I}_{J\;1\;J+1}\right)^2\left(N^{(1)}_{J'}\right)^2\right],\nonumber\\
&&\langle \phi^{(1)}_{IM}|H|\Psi^{(2qp;J1)}_{JI;M}\rangle =4X^{(\tau)}_{pc}\xi^{(+)}_{jj}N^{(2qp;J1)}_{JI}\delta_{J,2}\sum_{J'}
\left(N^{(1)}_{J'}\right)^{-1}C^{J'\;J\;I}_{1\;J;J+1}\langle \phi ^{(1)}_{I}||b^{\dag}_{\tau}+b_{\tau}||\phi^{(1)}_{J'}\rangle ,\tau=p,n\nonumber\\
&&\langle \Psi^{(2qp;J1)}_{JI;M}|H|\phi^{(1)}_{IM}\rangle=\langle \phi^{(1)}_{IM}|H|\Psi^{(2qp;J1)}_{JI;M}\rangle.
\end{eqnarray}
The notation W(abcd;ef) stands for the Racah coefficients.  The isospin quantum number $\tau$ takes the values $p$ or $n$ depending on whether the  two quasiparticle component is of proton or of neutron nature and, moreover, the model Hamiltonian describes the coupling of the $\tau$-like particles to the core.

We note that the matrix elements of the model Hamiltonian are expressed in terms of the reduced matrix elements of the quadrupole operators between states belonging to the  phenomenological dipole band. These are given analytically below:
\begin{eqnarray}
\langle\phi^{(1)}_{I^{\prime}}||b_{\tau}||\phi^{(1)}_{I}\rangle &=&d\frac{2I+1}{2I^{\prime}+1}C^{I\;2\;I^{\prime}}_{1\;0\;1}\frac{N^{(1)}_{I}}{N^{(1)}_{I^{\prime}}}
                                                                +3d\hat{I}N^{(1)}_{I}N^{(1)}_{I^{\prime}}\sum_{I_1I_2}F^{I^{\prime}I}_{I_1I_2}C^{I_1\;1\;I^{\prime}}_{0\;1\;1}
                                                                  \left(N^{(g)}_{I_1}\right)^{-2},\nonumber\\  
F^{I^{\prime}I}_{I_1I_2}&=&\hat{I}_2C^{2\;1\;I_2}_{0\;1\;1}C^{I_1\;I_2\;I^{\prime}}_{0\;1\;1}W(2 2 I_2 2; 1 1)W(I^{\prime} 2 I_1 I_2; I 1),\nonumber\\
\langle\phi^{(1)}_{I}||b^{\dagger}_{\tau}||\phi^{(1)}_{I^{\prime}}\rangle &=&\frac{\hat{I}^{\prime}}{\hat{I}}(-1)^{I-I^{\prime}}\langle\phi^{(1)}_{I^{\prime}}||b_{\tau}||\phi^{(1)}_{I}\rangle ,
 \tau=p,n . 
\end{eqnarray}


\begin{references}
\bibitem{LoIu1} N. Lo Iudice and F. Palumbo, Phys. Rev. Lett. {\bf 41}, 1532 (1978).
\bibitem{LoIu2} G. De Francheschi, F. Palumbo and N. Lo Iudice, Phys. Rev. {\bf C29} (1984) 1496.
\bibitem{Zawischa} D. Zawischa, J. Phys. G{\bf 24}, 683,(1998).
\bibitem{LoIu3} N. Lo Iudice, Phys. Part. Nucl. {\bf 25 }, 556, (1997).
\bibitem{Frau}S. Frauendorf, Rev. Mod. Phys.  {\bf 73} (2001) 463.
\bibitem{Jenkins} Jenkins et al., Phys. Rev. Lett. {\bf 83} (1999) 500.
\bibitem{1}A. A. Raduta, A. Faessler and V. Ceausescu, Phys. Rev. 
{\bf C36} (1987) 2111.
\bibitem{2} A. A. Raduta, I. I. Ursu and D. S. Delion, Nucl. Phys. 
{\bf A 475} (1987) 439.
\bibitem{3} A. A. Raduta and D. S. Delion, Nucl. Phys. {\bf A 491} (1989) 24.
\bibitem{4} N. Lo Iudice, A. A. Raduta and D. S. Delion, Phys. Lett. 
{\bf B 300} (1993) 195; Phys. Rev. {\bf C 50} (1994) 127.
\bibitem{5}A. A. Raduta, D.S. Delion and N. Lo Iudice, Nucl. Phys. 
{\bf A564} (1993) 185.
\bibitem{6} A. A. Raduta, I. I. Ursu and Amand Faessler, Nucl. Phys. {\bf A 489} (1988) 20.
\bibitem{7}A. A. Raduta, A. Escuderos and E. Moya de Guerra, Phys. Rev. 
{\bf C 65} (2002) 0243121. 
\bibitem{8} A. A. Raduta, N. Lo Iudice and I. I. Ursu, Nucl. Phys. {\bf 584} (1995) 84.
\bibitem{9} A. A. Raduta, Phys. Rev {C \bf A51} (1995) 2973.
\bibitem{10} A. Aroua, {\it et al}, Nucl. Phys. {\bf A728} (2003) 96.
\bibitem{11}A. A. Raduta, C.M. Raduta and Amand Faessler, Phys. Lett. B, 635 (2006) 80.
\bibitem{12}A. A. Raduta, Al. H. Raduta and C. M. Raduta,  Phys. Rev. C74 (2006) 044312.
\bibitem{13} Raduta et al., Phys. Rev. C 80, 044327 (2009).
\bibitem{Rad1}A. A. Raduta, V. Ceausescu, A. Gheorghe and R. Dreizler, Phys. Lett. {\bf B 99} (1981) 444; Nucl. Phys. {\bf A 381}
(1982) 253.
\bibitem{Rad2}A. A. Raduta, A. Faessler and V. Ceausescu, Phys. Rev. {\bf C 36} (1987) 439.
\bibitem{Rad3}A. A. Raduta, I. I. Ursu and D. S. Delion, Nucl. Phys. {\bf A 475} (1987) 439.
\bibitem{Rad4}A. A. Raduta and D. S. Delion, Nucl. Phys. {\bf A 491} (1989) 24.
\bibitem{Iud}N. Lo Iudice, A. A. Raduta and D. S. Delion, Phys. Rev. {\bf C50} (1994) 127.
\bibitem{Lima}A. A. Raduta, C. lima and Amand Faessler, Z. Phys. A - Atoms and Nuclei {\bf 313}, (1983), 69.
\bibitem{Coral} Coral M. Baglin, Nuclear Data Sheets 113 (2012) 1871.
\bibitem{Petrache96}C. M. Petrache {\it et al.,} Nucl. Phys. {\bf A597} (1996) 106.
\bibitem{Simon2005}A. j. Simon {\it et al.,} Jour. Phys. G: Nucl. Part. Phys {\bf 31} (2005) 541.
\bibitem{Vaman}C. Vaman, D. B. Fossan, T. Koike, K. Starosta, I. Y. Lee and A. O. Machiavelli, Phys. Rev. Lett. {\bf 92} (2004) 032501.
\bibitem{Balab} D. L. Balabanski, {\it et al.,} Phys. Rev. C {\bf 70} (2004) 044305.
\bibitem{Frau97}S. Frauendorf and J. Meng, Nucl. Phys. {\bf A 617} (1997) 131.
\bibitem{Fas1} H. Toki and Amand Faessler, Nucl. Phys. {\bf A 253} (1975) 231.
\bibitem{Fas2} H. Toki and Amand Faessler, Z. Phys.. {\bf A 276} (1976) 35.
\bibitem{Fas3} H. Toki and Amand Faessler, Phys. Lett. {\bf B 63} (1976) 121.
\bibitem{Fas4} H. Toki, H. L. Yadav and Amand Faessler, Phys. Lett. {\bf B 66} (1977) 310.
\bibitem{Petrache06}C. M. Petrache, G. B. Hegemann, I. Hamamoto and K. Starosta, Phys. Rev. Lett. {\bf 96} (2006) 112502.
\bibitem{Rad73}A. A. Raduta, V. Ceausescu, Gh. Stratan and A. Sandulescu, Phys. Rev {\bf C 8} (1973) 1525.
\bibitem{Rad74} V. Ceausescu and A. A. Raduta, Prog. Th. Phys. {\bf 52} (1974) 903.
\bibitem{RadRadRad} A. A. Raduta, Al. H. Raduta and c. M. Raduta, Phys. Rev. {\bf C74} (2006) 044312.
\bibitem{Robledo}L. M. Robledo, Phys. Rev. C {\bf 50}, 2874 (1994).
\bibitem{Egido1} J. L. Egido, L. M. Robledo, and Y. Sun, Nucl. Phys. A {\bf 560}, 253 (1993).
\bibitem{Guzman} R. Rodrigues-Guzman, J. L. Egido, and L. M. Robledo, Nucl. Phys. A {\bf 709}, 201 (2002).
\bibitem{Egido2} J. L. Egido and L. M. Robledo, arhiv: nucl-th/0311106v1, 28 Nov. 2003;Lect. Notes Phys. {\bf 641}, 269 (2004).
\end{references}
\end{document}